\documentclass[]{interact}

\usepackage{mathrsfs}  

\usepackage{color}
\usepackage{tikz,pgfplots,tikz-3dplot}
\pgfplotsset{compat=newest}
\usetikzlibrary{arrows,decorations.markings,shapes,shapes.multipart,arrows.meta}
\usetikzlibrary{chains}
\usepackage{hyperref}
\usepackage{standalone}

\newcommand{\aver}[1]{\! \left\langle {#1} \right\rangle \!}
\newcommand{\vect}[1]{\bm{#1}}
\newcommand{\rmd}{\text{d}}

\begin{document}

\title{An efficient numerical method \\ for the Generalised Kolmogorov Equation}
\author{
\name{D. Gatti\textsuperscript{a}, 
A. Remigi\textsuperscript{b}, 
A. Chiarini\textsuperscript{b}, 
A. Cimarelli\textsuperscript{c} and 
M. Quadrio\textsuperscript{b}}
\affil{
\textsuperscript{a}Institute of Fluid Mechanics, Karlsruhe Institute of Technology, Germany;\\
\textsuperscript{b}Dipartimento di Scienze e Tecnologie Aerospaziali, Politecnico di Milano, Italy;\\
\textsuperscript{c}Dipartimento di Ingegneria Industriale e Scienze Matematiche, Universit\'a Politecnica delle
Marche, Italy}
}

\maketitle

\begin{abstract}
An efficient algorithm for computing the terms appearing in the Generalised Kolmogorov Equation (GKE) written for the indefinite plane channel flow is presented. The algorithm, which features three distinct strategies for parallel computing, is designed such that CPU and memory requirements are kept to a minimum, so that high-$Re$ wall-bounded flows can be afforded.

Computational efficiency is mainly achieved by leveraging the Parseval's theorem for the two homogeneous directions available in the plane channel geometry. A speed-up of 3-4 orders of magnitude, depending on the problem size, is reported in comparison to a key implementation used in the literature. Validation of the code is demonstrated by computing the residual of the GKE, and example results are presented for channel flows at $Re_\tau=200$ and $Re_\tau=1000$, where for the first time they are observed in the whole four-dimensional domain. It is shown that the space and scale properties of the scale energy fluxes change for increasing values of the Reynolds number. Among all scale energy fluxes, the wall-normal flux is found to show the richest behaviour for increasing streamwise scales.
\end{abstract}


\clearpage{}\section{Introduction} 

Characterising a turbulent flow from the energetic standpoint has always been an important endeavour in turbulence research: laminar and turbulent flow regimes possess different energy requirements, and such distinction often becomes of paramount importance in applications.  

The seminal contributions by Richardson \cite{richardson-1922} and Kolmogorov \cite{kolmogorov-1941} described how, in the idealised setting of an homogeneous and isotropic flow, turbulent energy is distributed within a hierarchy of eddies, characterised by different length scales; the concept of energy cascade was introduced, which is understood as a flux of energy across scales. Such a description also applies to inhomogeneous flows, but in this case an additional spatial redistribution of energy occurs, giving rise to a spatial energy flux. In the geometrically simple setting of an indefinite plane channel, where the wall-parallel directions are statistically homogeneous, the spatial flux takes place in the wall-normal direction only. Hence, in wall-turbulence two types of energy fluxes coexist: one is best described in the space of scales (eddy size), while the other is naturally observed in physical space (wall-distance). 

In the last 30 years, also thanks to the comprehensive information available from Direct Numerical Simulations (DNS) of turbulent wall flows in canonical geometries, the structure of a wall-bounded flow in the wall-normal direction has been thoroughly studied; the flow domain is subdivided into several regions where different phenomena contribute to the energy budget with different relative importance. Such analysis is nowadays the accepted way of describing energy exchanges in wall-bounded turbulent flows \cite{kim-moin-moser-1987, mansour-kim-moin-1988}, although it does not provide yet an entirely satisfactory description; one of the reasons lies in the missing link with the energy cascade concept. 

A complementary approach capable to provide an unified view of the energetics of turbulent flows, is progressively gaining acceptance in the recent years. It hinges upon a differential equation that, in its original form valid for homogeneous and isotropic turbulence, can be traced back to Kolmogorov himself \cite{kolmogorov-1941, kolmogorov-dissipation-1941}. The so-called Kolmogorov equation, counterpart of the K\'arm\'an-Howarth equation for the correlation function, is an exact transport equation for the second-order structure function, i.e. the variance of velocity differences between two points $\vect{x}$ and $\vect{x'}$ in the fluid domain. The Kolmogorov equation has been generalised to inhomogeneous anisotropic flows by Hill \cite{hill-2001, hill-2002}, thus paving the way towards a unified description of energetically relevant phenomena in both physical and scale spaces.

This extended form, known as the Generalised Kolmogorov Equation (GKE), has been used in several papers to study the energetics of inhomogeneous flows in the complete space of scales and positions. For the sake of brevity, we mention here only few of them. A generalised form of the Kolmogorov equation was studied in Danaila et al. \cite{danaila-etal-2001} to address the influence of the spatial inhomogeneity of the large scales on the energy budget of a turbulent channel. In the same geometry and exploiting DNS data, Marati et al. \cite{marati-casciola-piva-2004} used the GKE to systematically characterise for the first time the spatial flux and energy cascade processes of the different flow regions of wall-turbulence. This approach was further developed by Cimarelli et al. \cite{cimarelli-deangelis-casciola-2013, cimarelli-etal-2016} who also discussed turbulence modelling issues \cite{cimarelli-deangelis-2011, cimarelli-deangelis-2012}. The GKE terms have been computed also in \cite{gomes-ganapathisubramani-vassilicos-2015} by using particle image velocimetry measurements, and in \cite{portela-papadakis-vassilicos-2017} by using DNS data, to study the most inhomogeneous and anisotropic regions in the wake of a grid-generated turbulence and of a square prism, respectively. The inhomogeneous development of the scale-by-scale budgets in a turbulent round jet was studied in Burattini et al. \cite{burattini-antonia-danaila-2005}, while \cite{thiesset-antonia-danaila-2013} addressed the intermediate wake of two-dimensional wake generators. Finally, the Kolmogorov equation has been also studied in thermally-driven turbulent flows together with its counterpart equation for the scalar field, the so-called Yaglom equation, in Danaila et al. \cite{danaila-etal-2012} and Gauding et al. \cite{gauding-etal-2014}; Togni et al. \cite{togni-cimarelli-deangelis-2015} for the first time employed the exact equations to study Rayleigh--B\'enard convection.

A seldom discussed but key feature of the GKE is the extent of its computational requirements. The GKE is an equation for the second-order structure function, which in the most general case depends upon 6 independent variables, the coordinates of the two points $\vect{x}$ and $\vect{x'}$. 
This is the fundamental reason why computing the GKE terms is so challenging. From an experimental point of view, it is difficult to simultaneously access the three-dimensional velocity and pressure fields at two points while spanning the whole flow domain. When such information is available, e.g. when processing a DNS dataset, the size of the computational problem associated with the evaluation of the GKE terms becomes huge, with obvious consequences in terms of both computing time and memory requirements. Indeed, the number of operations required to compute every term of the GKE is of the order of $N_t N^2$ where $N_t$ is the number of flow snapshots available for time average, and $N$ is the total number of points used to discretise the flow domain. Since $N$ is in excess of one million even for a basic DNS in a plane channel flow at low values of the Reynolds number, and quickly increases as $Re$ increases, one must be aware that computing the GKE terms may require way more computational effort than creating the DNS database itself. The availability of an efficient and memory-friendly GKE computer code is essential to address high-$Re$ flows, which bring about computational problems of rapidly growing size.

A further difficulty posed by the GKE lies in the graphical representation of the 6-dimensional compound space of scales and positions. Even for the simplified case of the plane channel, which possesses two homogeneous directions, the independent variables are four (the separations in the three spatial directions, and the wall-normal position), and dealing with variables defined in a 4-dimensional space remains quite complex. Indeed, in the first paper where the terms of the GKE were actually computed for a channel flow \cite{marati-casciola-piva-2004}, further assumptions had to be made, and the GKE was integrated over a square plane of edge $r$ in a wall-parallel plane, under the assumption of zero wall-normal separation. The simplifications made in \cite{marati-casciola-piva-2004} reduced the independent variables down to two (the square edge $r$ and the wall-normal coordinate). Over the years, the analysis of the GKE was further developed and refined: Cimarelli et al. in \cite{cimarelli-deangelis-casciola-2013} and, more recently, in \cite{cimarelli-etal-2016} extended the analysis to two different 3-dimensional subspaces, whereas \cite{mollicone-etal-2018} considered different 2-dimensional subspaces of the 5-dimensional GKE defined for a streamwise-developing turbulent flow with separation.

Aim of the present paper is to describe an implementation of a new code for computing the terms of the GKE equation. The code, that is made available to the community with an open-source license, is tailored to the plane channel flow and is designed from scratch to be fast and efficient, both in terms of CPU and memory requirements. In fact, efficiency is key if one plans to observe how energy dynamics is modified as the Reynolds number increases. The implementation is properly tested and validated by using DNS databases produced {\em ad hoc}; the budget residual is examined to assess both the correctness of the implementation and the quality of the statistical convergence. The main design choices that make our implementation so much faster than the existing one(s) are discussed and motivated, and computing times are measured to report a speed-up that, for the problems tested, reached 4 orders of magnitude with respect to current implementations. Two channel flow cases at $Re_\tau=200$ and at $Re_\tau=1000$ are used to present example results and to analyse the effects of the Reynolds number in the multidimensional space of scales and positions; for the first time, the GKE terms are observed in the 4-dimensional space.

The structure of the paper is as follows. First in Sec.\ref{sec:gke} the second-order structure function and the GKE in its specialised form tailored to a plane channel flow are briefly introduced. Then in Sec.\ref{sec:algorithm} the implementation of our code is described in detail, together with the main design choices and the parallel strategies. Finally, in Sec.\ref{sec:results} the performance of the new implementation is discussed, and  in Sec.\ref{sec:physics} for the first time an analysis of the GKE in the complete 4-dimensional space is provided.  
\clearpage{}
\clearpage{}\section{The Generalised Kolmogorov equation}
\label{sec:gke}

We consider an indefinite plane channel flow, with a Cartesian coordinate system in which $x$ and $z$ denote the homogeneous streamwise and spanwise directions respectively, whereas $y$ is the wall-normal direction. The corresponding velocity components are $\tilde{u}$, $\tilde{w}$ and $\tilde{v}$. The index notation $x_i$, $\tilde{u}_i$ is also used, with $i=1,3$ identifying the homogeneous directions and $i=2$ the wall-normal one. The Reynolds decomposition is used in such a way that upper- and lower-case symbols denote mean and fluctuating quantities, i.e. $\tilde{u}_i = U_i + u_i$ and $\tilde{p} = P + p$. The two parallel walls are separated by a gap of width $2h$, and the Reynolds number for the problem is defined by using $h$ as the reference length scale. If the friction velocity $u_\tau$ is used as the velocity scale, the Reynolds number becomes the so-called friction Reynolds number:
\[
Re_\tau = \frac{u_\tau h}{\nu},
\]
with $\nu$ the kinematic viscosity of the fluid. 

\begin{figure}
\centering
 \includegraphics[]{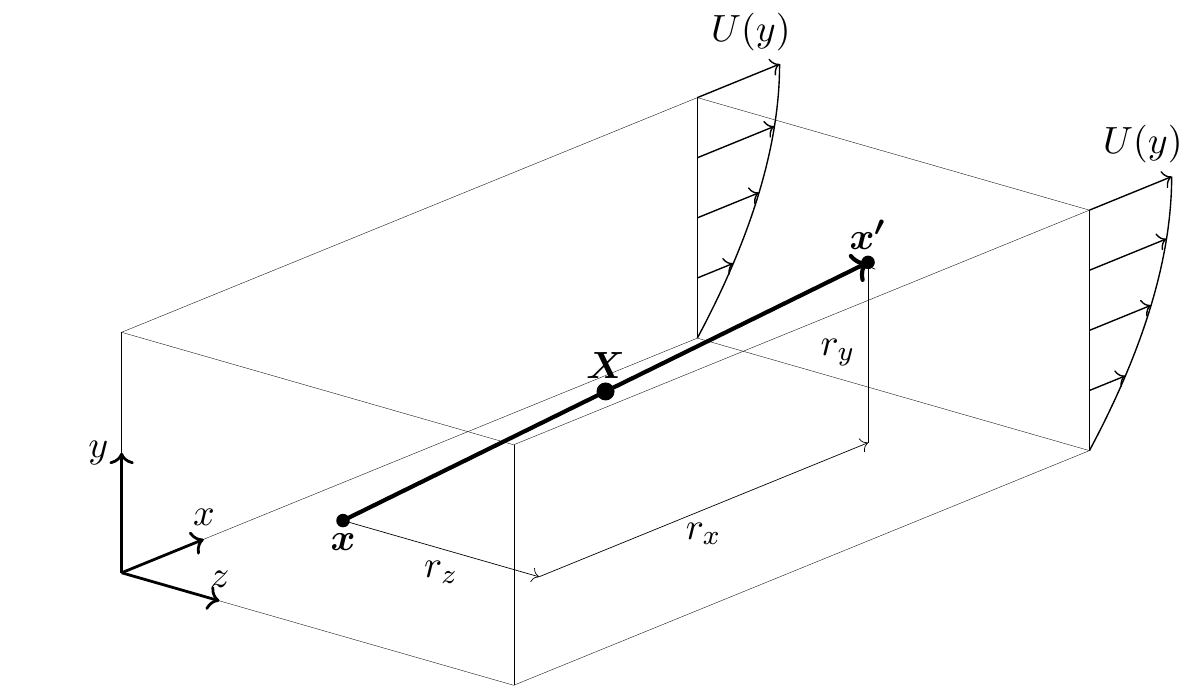}
 \caption{Definition of the structure function $\aver{\delta u^2}$, variance of the velocity difference between the two points $\vect{x}$ and $\vect{x'}$ with separation vector $\vect{r}$, in the geometry of the plane channel.}
\label{fig:coord_scheme}
\end{figure}

Let $\vect{x}=\vect{X} - \vect{r}/2$ and $\vect{x}'=\vect{X} + \vect{r}/2$ be two points inside the fluid domain, separated by a separation vector $\vect{r}$ as exemplified in fig.\ref{fig:coord_scheme}, and let $\vect{X}$ be the mid-point. The quantity $\delta \vect{u} = \vect{u}' - \vect{u}$ is the difference between the two velocity vectors evaluated at $\vect{x}'$ and $\vect{x}$. 

The GKE is a budget equation in conservative form for the second-order structure function $\aver{\delta u^2}$, defined as the variance of the velocity difference
\begin{equation}
\langle \delta u^2 \rangle = \langle \delta u_i \delta u_i\rangle \,
\label{du2}
\end{equation}
where repeated indices imply summation, and angular brackets denote ensemble average as well as averaging along homogeneous directions, if available, and over time, if the flow is statistically stationary. $\aver{\delta u^2}$ is considered \cite{davidson-2004} to represent the energy associated with the scales of motion up to $\vect{r}$, and for that reason is here referred simply to as scale energy. Such interpretation is phenomenologically consistent with the observation that any eddy of size $L \gg \left| \vect{r} \right| $ induce a negligible velocity different at a separation $\vect{r}$. Literally, the GKE provides an exact balance equation between second- and third-order moments of a central quantity in turbulence studies, that is the velocity increment $\delta u_i$ \cite{germano-2007}. Furthermore, the GKE is also an exact equation for the rate of dissipation of turbulent kinetic energy, $\epsilon$, which is associated with third-order moments of the velocity increment at every scale and position. Only in this context, the second- and third-order moments assume their physical interpretation of scale energy and scale-energy fluxes.

In general, $\aver{\delta u^2}$ is function of the separation vector $\vect{r}$ and the mid-point $\vect{X}$, i.e. function of 6 independent variables. An exact equation for $\aver{\delta u^2}$ was derived first by Kolmogorov in \cite{kolmogorov-dissipation-1941} for isotropic turbulence, and has been later generalised to inhomogeneous flows. We follow Cimarelli et al. \cite{cimarelli-etal-2016} and write below the GKE in the specialised form valid for the indefinite plane channel flow. This form of the GKE was previously introduced in \cite{marati-casciola-piva-2004}, and later refined with the addition of a couple missing terms. In the case of the plane channel flow, the number of independent variables reduces to 4, as for the mid-point $\vect{X}$ only its wall-normal coordinate $Y = X_2$ matters. Moreover, the mean flow possesses only one non-zero component with the present choice of coordinate system; this component only varies along the wall-normal coordinate, i.e. $\aver{\vect{U}}=(U(y),0,0)$. Thus, the GKE for plane channel flow reads:
\begin{equation}
\begin{split}
\frac{\partial \aver{\delta u^2 \delta u_i}}{\partial r_i} + 
\frac{\partial \aver{\delta u^2 \delta U}}{\partial r_x} 
- 2 \nu \frac{ \partial^2 \aver{\delta u^2} }{\partial r_i \partial r_i}
+ \frac{\partial \aver{v^* \delta u^2}}{\partial Y}
+ \frac{2}{\rho} \frac{\partial \aver{\delta p \delta v}}{\partial Y}
- \frac{\nu}{2}  \frac{\partial^2 \aver{\delta u^2}}{\partial Y^2} = \\
= - 2 \aver{\delta u \delta v} \left( \frac{\rmd U}{\rmd y} \right)^*
- 2 \aver{\delta u v^*} \delta \left( \frac{\rmd U}{\rmd y} \right)
-4 \aver{\epsilon^*}.
\end{split}
\label{eq:GKE}
\end{equation}

In this expression, the asterisk denotes the arithmetic average of a variable evaluated at $\vect{x}$ and $\vect{x}'$, $\nu$ is the kinematic viscosity and $\epsilon= \nu \left( \partial u_i / \partial x_j \right) \left( \partial u_i /\partial x_j \right)$ is the pseudo-dissipation rate of turbulent kinetic energy. The GKE contains source terms, written at the r.h.s., which act as production or dissipation depending on their sign. Since the l.h.s. can be written as divergence of fluxes, the equation written in conservative form becomes:
\begin{equation}
\vect{\nabla_r} \cdot \vect{\Phi} \left( Y,\vect{r} \right) + \frac{\partial \phi \left( Y,\vect{r} \right)}{\partial Y}
= \xi \left( Y, \vect{r}\right).
\label{eq:GKE-conservative}
\end{equation}

In Eq.\eqref{eq:GKE-conservative}, $\vect{\Phi}$ and $\phi$ are the flux in the space of scales and the flux in the physical space, whereas $\xi$ is the source term. The operator $\vect{\nabla_r}$ is the gradient operator in the $\vect{r}$ space. By comparing Eq.\eqref{eq:GKE-conservative} with Eq.\eqref{eq:GKE} one easily arrives at the definitions of the fluxes and of the source term. The scale-energy flux vector $\vect{\Phi}$ is
\begin{equation}
\vect{\Phi} \left( Y, \vect{r} \right) = 
\aver{\delta u^2 \delta \vect{u}}
- 2 \nu \vect{\nabla_r} \aver{\delta u^2}
+ \aver{\delta u^2 \delta U} \vect{e}_x
\label{eq:scale-flux}
\end{equation}
where $\vect{e}_x$ is the unit vector in the streamwise direction. The spatial flux of scale energy $\phi$ is
\begin{equation}
\phi \left( Y, \vect{r} \right) = 
\aver{v^* \delta u^2} 
+ \frac{2}{\rho} \aver{\delta p \delta v} 
- \frac{\nu}{2} \frac{\partial \aver{\delta u^2}}{\partial Y} .
\label{eq:physical-flux}
\end{equation}
In both fluxes, viscous terms are recognised, which quantify viscous transport of scale energy, and turbulent terms, which quantify turbulent transport of scale energy. In the latter, the second-order structure function is coupled with turbulent fluctuations. Moreover, one term in $\phi$ accounts for pressure-velocity coupling, and another one in $\vect{\Phi}$ accounts for the coupling with the mean flow. Finally, the scale-energy source is
\begin{equation}
\xi\left(Y, \vect{r} \right) =
- 2 \aver{\delta u \delta v} \left( \frac{\rmd U}{\rmd y} \right)^*
- 2 \aver{\delta u v^*} \delta \left( \frac{\rmd U}{\rmd y} \right)
- 4 \aver{\epsilon^*} ,
\label{eq:source}
\end{equation}
and the flow regions with $\xi>0$ (production larger than dissipation) are those where the scale energy is produced.

As for every conservation law, also in the GKE fluxes are defined up to an arbitrary solenoidal field, as demonstrated for example by Jim\'enez \cite{jimenez-2016}. In other words, the fluxes of scale energy as in the above definitions are not uniquely defined, since they are obtained from a manipulation of the governing equations which leads to a specific form of the GKE. The present form is selected as it carries a direct correspondence with the more familiar form of the single-point turbulent kinetic energy equation.

\clearpage{}
\clearpage{}\section{Structure of the GKE computer code}
\label{sec:algorithm}

In this section the implementation of an efficient strategy for computing the budget of the Generalised Kolmogorov Equation tailored to the indefinite plane channel flow, i.e. Eq.\eqref{eq:GKE}, is discussed. The code only inherits a small set of design choices from the accompanying DNS code, described in \cite{luchini-quadrio-2006}, mainly the programming language and the type of spatial discretisation. Hence, its organisation, designed to minimise CPU and memory requirements, carries a general interest. The source code, which is entirely self-contained, is freely available via GitHub at this \href{https://github.com/davecats/gke}{link}. The source code is quite short (about 100 lines for the core part) and having it at hand can be helpful to understand the code structure described below.

\begin{figure}
 \centering
 \includegraphics[]{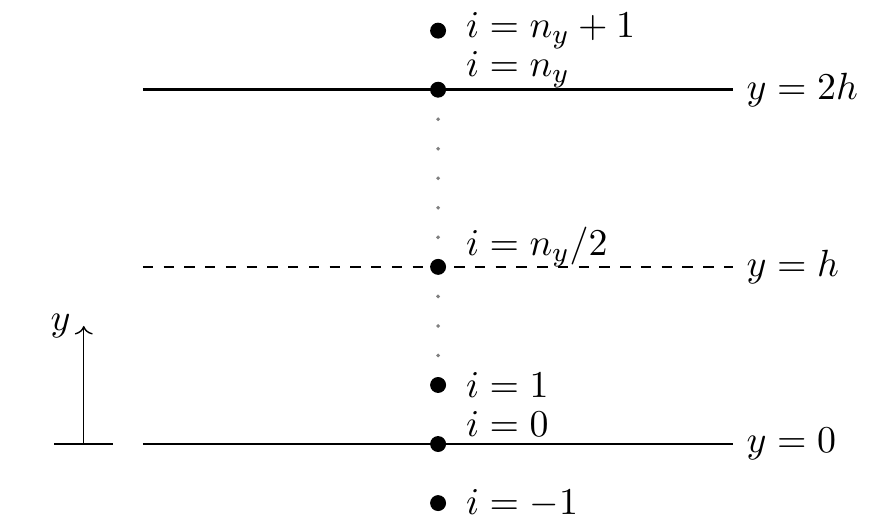}
 \caption{Sketch of the grid points in the wall-normal direction: the indices $i=0$ and $i=n_y$ identify the grid points at the two walls, $y=0$ and $y=2h$. The grid possesses two ghost nodes for $i=-1$ and $i=n_y+1$.}
\label{fig:grid-points}
\end{figure}

Computing the budget of the GKE is typically a post-processing step which operates on a database previously created by a DNS code. Irrespective of the discretisation strategy adopted in the DNS one can generally assume that the database is composed of (or can be translated into) a number $N_t$ of temporal snapshots of velocity fields. Every snapshot obeys the set of boundary conditions of the differential problem (periodicity at the inflow/outflow, no-slip and no-penetration at the solid walls) and is available with velocity values known in a collocated form at every point of a Cartesian grid. In particular, we assume data are available on a finite set of wall-normal positions, denoted as $y_i$, with $-1 \le i \le n_y+1$. The two walls are located at $y=0$ and $y=2h$, corresponding to $i=0$ and $i=n_y$. The values $i=-1$ and $i=n_y+1$ denote ghost nodes used to compute $y$-derivatives at the walls (see figure \ref{fig:grid-points}). The specific structure of the databases produced and used in this paper is further discussed in Sec.\ref{sec:results}, where results are presented.

The main characteristics of the code are: (i) all the GKE terms, i.e. $\Phi_{r_x}$, $\Phi_{r_y}$, $\Phi_{r_z}$, $\phi$, $\xi$ and $\aver{\delta u^2}$, are rewritten in a form that involves multiple but simpler correlations: for the homogeneous directions, the Parseval theorem is then used to compute them in Fourier space, with huge computational advantage (see \S\ref{sec:homogeneous_directions}); (ii) the GKE terms, computed in a four-dimensional domain, depend on the two independent variables $Y$ and $r_y$; however, in computing them we switch to the equivalent variables $Y_1$ and $Y_2$, defined as $Y_1=Y-{r_y}/2$ and $Y_2=Y+{r_y}/2$. This simplifies taking advantage of their symmetries (see \S\ref{sec:symmetries}) to further reduce the computational effort and the amount of memory required by the code.

\subsection{Products instead of correlations}
\label{sec:homogeneous_directions}

When homogeneous directions are available, as in the present case of the indefinite plane channel, the GKE terms can be rewritten in a different form, which allows computing them in a simpler and computationally efficient way. As an example, we focus in the following on a specific term, i.e. $\aver{\delta u \delta v}$ appearing in the definition \eqref{eq:source} of the source $\xi$, but the same reasoning holds true in general for all terms appearing in the definition of the GKE. With simple algebra the term $\aver{\delta u \delta v}$ can be rewritten in terms of the two-point correlation
\begin{equation*}
\aver{uv}(r_x,r_z;Y_1,Y_2)=\aver{u(x,Y_1,z)v(x+r_x,Y_2,z+r_z)} \, ,
\end{equation*}
as follows
\begin{equation}
\begin{split}
\aver{\delta u \delta v}(r_x,r_y,r_z,Y)=&
- \aver{uv}(r_x,r_z;Y_1,Y_2) - \aver{vu}(r_x,r_z;Y_1,Y_2)+ \\
&+ \aver{uv}(0,0;Y_1,Y_1) + \aver{uv}(0,0;Y_2,Y_2) \, .
\end{split}
\label{eq:expansion}
\end{equation}
The notation emphasises that we are concerned with the homogeneous directions only.

Equation \eqref{eq:expansion} above transforms $\aver{\delta u \delta v}$ into a sum of four correlations. As the present problem enjoys two homogeneous directions, for which a representation in Fourier space is always possible and indeed very popular in the DNS practice, correlations in Fourier space can be advantageously computed by resorting to the Parseval theorem, thus achieving the same computational efficiency that lies at the root of the pseudo-spectral method for the DNS of incompressible channel flow.

The GKE terms are thus not computed directly, but assembled after computing in Fourier space the required set of cross-correlations. If again the term $\aver{uv}(r_x,r_z;Y_1,Y_2)$ in Eq.\eqref{eq:expansion} is taken as an example, this two-points correlation is defined as (omitting for simplicity the temporal average): 
\begin{equation}
\aver{uv}(r_x,r_z;Y_1,Y_2) =
\int_{-\infty}^{+\infty} \int_{-\infty}^{+\infty} u(x,Y_1,z) v(x+r_x,Y_2,z+r_z) \text{d}x \text{d}z
\label{eq:correlations}
\end{equation}
and can be efficiently computed with a single product in Fourier space via the Parseval theorem. If $\hat{u}(Y_1)$ and $\hat{v}(Y_2)$ are the two-dimensional Fourier transforms of $u(x,z)$ and $v(x,z)$ respectively at planes $Y_1$ and $Y_2$, the Parseval theorem provides the following identity:
\begin{equation}
\int_{-\infty}^{+\infty} \int_{-\infty}^{+\infty} u(x,Y_1,z) v(x+r_x,Y_2,z+r_z) \text{d}x \text{d}z = \mathscr{F}^{-1}(\hat{u}^*(Y_1) \hat{v}(Y_2))
\end{equation}
where the $*$ superscript denotes the complex conjugate, and the operator $\mathscr{F}^{-1}$ is the inverse Fourier transform.

It must be mentioned that a few terms of Eq.\eqref{eq:GKE} contain triple correlations; one such example is
\begin{equation*}
\aver{uuv}(r_x,r_z;Y_1,Y_2)=\aver{u(x,Y_1,z)u(x,Y_1,z)v(x+r_x,Y_2,z+r_z)}
\end{equation*}
related to the scale flux vector $\vect{\Phi}$.
This term is computed as:
\begin{equation*}
\aver{uuv}(r_x,r_z;Y_1,Y_2)=\mathscr{F}^{-1}(\widehat{uu}^*(Y_1) \hat{v}(Y_2))
\end{equation*}
where $\widehat{uu}(Y_1)$ denotes an in-plane convolution in the Fourier space that can be efficiently computed as a product in the direct space:
\begin{equation*}
\widehat{uu}(Y_1)=\mathscr{F}(u(x,Y_1,z)u(x,Y_1,z)).
\end{equation*}

\subsection{General structure of the program}
\label{sec:structure}

\begin{figure}
 \centering
 \includegraphics[]{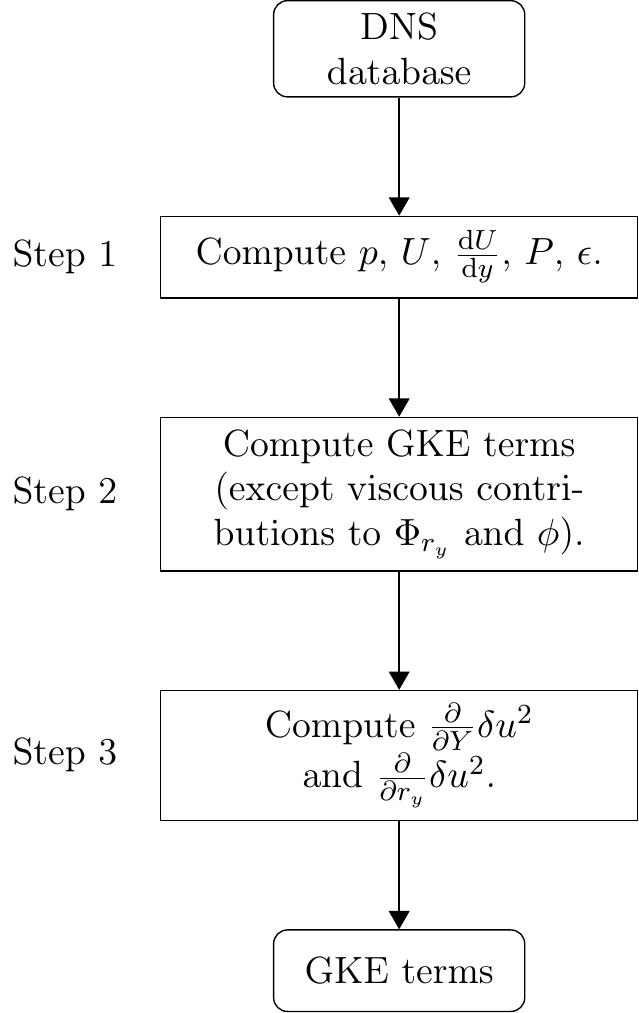}
 \caption{Steps leading from the DNS database to the GKE terms. The core calculations take place in Step 2.}
\label{fig:flowchart}
\end{figure}

The algorithm computes the terms of the GKE by analysing a DNS database composed of $N_t$ temporal snapshots. 

It consists in three steps (see figure \ref{fig:flowchart}) that are sequentially carried out to obtain the full set of GKE terms. In Step 1, all the required (e.g. mean) quantities are derived from the velocity fields. Then, in Step 2 the GKE terms are computed, by scanning the database once again and progressively averaging the contributions of every snapshot. Computing terms that require wall-normal derivatives, i.e. the viscous parts of the fluxes $\phi$ and $\Phi_{r_y}$, is deferred to Step 3, exploiting the commutative property of derivative and average operators to increase  efficiency. On the contrary, the viscous parts of $\Phi_{r_x}$ and $\Phi_{r_z}$ are separately computed for each snapshot and then averaged at the end of Step 2. Indeed, during Step 2 the algorithm allows under-sampling in the $r_x$ and $r_z$ directions, a feature that becomes progressively interesting as $Re$ increases. Since for maximum accuracy the derivatives of $\aver{\delta u^2}$ in the $r_x$ and $r_z$ directions are computed spectrally in the Fourier space, which would be unpractical when $\aver{\delta u^2}$ is known on a reduced grid, it is in general preferable to compute these fluxes in Step 2. Clearly, if undersampling is not used, computing the viscous parts of $\Phi_{r_x}$ and $\Phi_{r_z}$ too can be deferred to Step 3, thus increasing further the overall efficiency.

Hereafter the three steps of the algorithm are described in more detail.

\subsubsection{Step 1}
During Step 1, the instantaneous and mean quantities needed for the computations of the GKE terms are derived from the velocity fields. The pressure $p$, for example, is often neither required nor computed by the DNS solver, when the Navier--Stokes equations are solved in their velocity-vorticity form \cite{kim-moin-moser-1987}. Hence Step 1 is where the Poisson equation is solved to obtain the pressure field corresponding to each velocity field. Similarly, if the database only contains the wall-normal components of velocity and vorticity, this is where the full  velocity field is explicitly computed. Moreover, while scanning the whole set of $N_t$ temporal snapshots, the mean velocity profile $U$, the mean shear $\text{d}U/\text{d}y$, the mean pressure $P$ and the pseudo-dissipation $\epsilon$ profiles are computed.

\subsubsection{Step 2}

\begin{figure}
 \centering
 \includegraphics[]{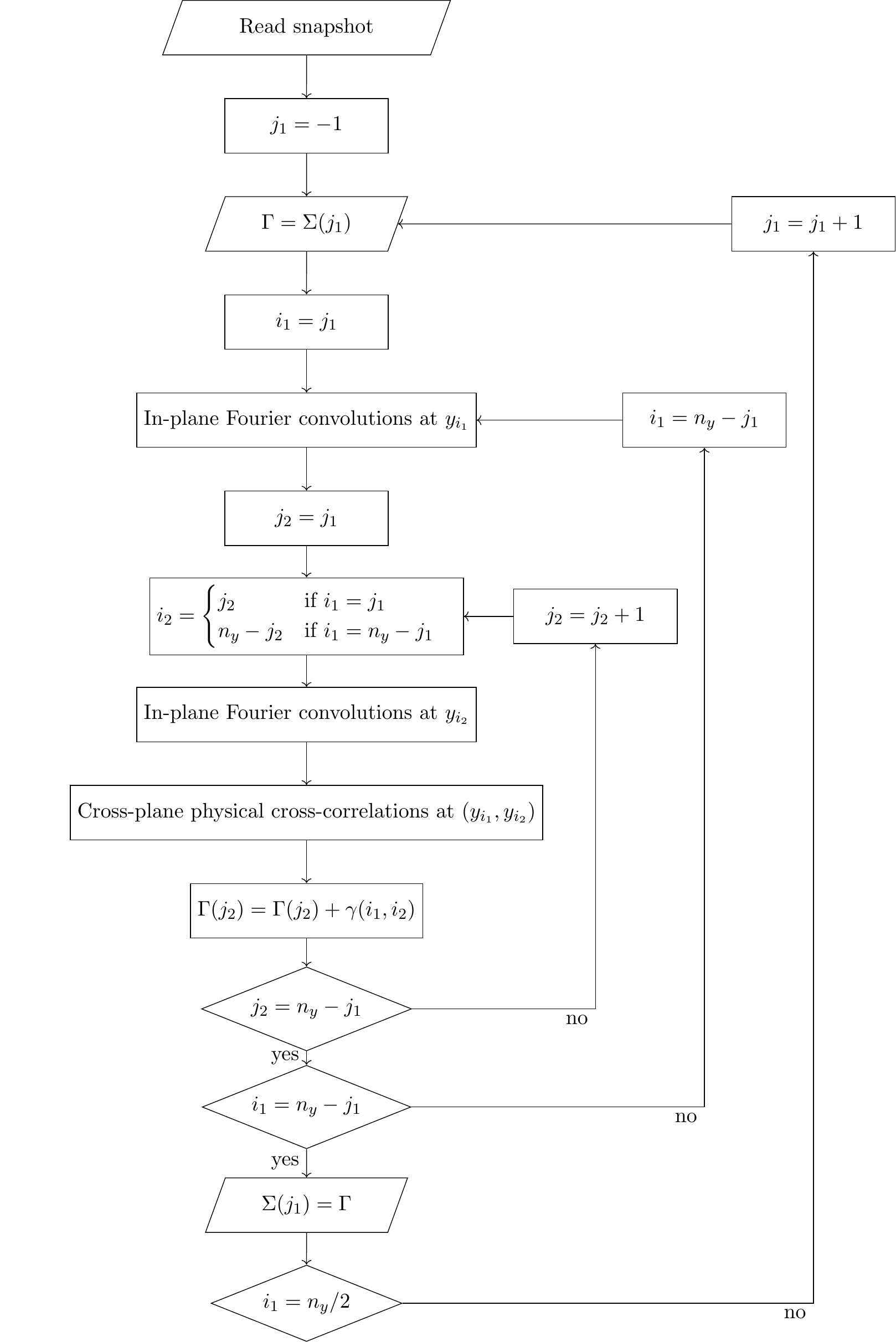}
 \caption{Structure of the core part of Step 2. The outer loop on $j_1$ spans half the set of wall-normal positions to double the statistical sample. The inner loop on $j_2$ exploits symmetries to minimise computing requirements.}
\label{fig:flowchart-step2}
\end{figure}

Step 2 is the core of the algorithm, where most of the GKE terms are computed. Its structure is illustrated by the flowchart in figure \ref{fig:flowchart-step2}, where for simplicity the average over the different snapshots is omitted. Within this Step, the data structure $\Sigma(Y_1)$ resides on disk and contains the full set of the GKE terms at position $Y_1$, and the structure $\Gamma(Y_1)$ is a memory array where $\Sigma$ is stored. Finally, $\gamma(Y_1,Y_2)$ contains contribution to the GKE terms from the pair $(Y_1,Y_2)$. 
 
Step 2 is made by two main nested loops. At the outer level, the code loops on $Y_1$; at the inner level, on $Y_2$. Since all the GKE terms are either symmetric or anti-symmetric with respect to an inversion of the wall-normal axis, one half of the channel is used to increase the size of the statistical sample. Hence, $Y_1$ scans through half the grid points in the wall-normal direction, i.e. loops from $y_{-1}$ to $y_{{n_y}/2}$, but for each $(Y_1,Y_2)$ also the terms from the pair $(2h-Y_1,2h-Y_2)$ are computed to contribute to the statistics, with the sign of each term properly set according to the relevant symmetry. For each $Y_1$, the GKE terms are computed for $Y_2$ ranging from $Y_1$ to $2h-Y_1$. As shown below in Sec.\ref{sec:symmetries}, when $Y_2<Y_1$ or $Y_2>2h-Y_1$ the GKE terms for the pair $(Y_1,Y_2)$ are computed from available information by exploiting symmetries.

In the flowchart of figure \ref{fig:flowchart-step2}, the indices $i_1$ and $j_1$ are used to select the wall-normal position $Y_1$, and the indices $i_2$ and $j_2$ to select $Y_2$. The index $i$ identifies the pair $(Y_1=y_{i_1},Y_2=y_{i_2})$ at which the GKE terms are actually evaluated; the index $j$ identifies the corresponding pair $(Y_1=y_{j_1},Y_2=y_{j_2})$ at which those terms are actually used. It is $i=j$ when $y_{i_1}<h$, and $i=n_y-j$ when $y_{i_1}>h$.

The sequence of operations performed by Step 2 for each snapshot is as follows. First, the entire snapshot is read and copied in memory. In the outer loop, for each $y_{j_1}$ the GKE terms for this position, i.e. $\Sigma(j_1)$, already computed and stored on disk while processing the previous velocity field, are read and copied into the memory array $\Gamma$. Then, in-plane Fourier convolutions terms at $y_{i_1}$, as for example $\widehat{uu}(y_{i_1})$, are computed in physical space. Now, in the inner loop on $j_2$, for each pair $(y_{i_1},y_{i_2})$ the contribution $\gamma$ to the GKE terms at $(y_{j_1},y_{j_2})$ is computed; first, the in-plane Fourier convolutions at $y_{i_2}$ are evaluated in physical space; and then the required cross-plane correlations, as for example  $\aver{uv}(r_x,r_z;y_{i_1},y_{i_2})$, are computed in Fourier space; lastly, the correlations are assembled and added to $\Gamma$. To double the size of the statistical sample, this set of computations is performed twice: for $i=j$ and $i=n_y-j$

After the inner $j_2$ loop is complete, $\Gamma$ is eventually written to disk as the updated $\Sigma(j_1)$. At the very end, i.e. when the last temporal snapshot is reached, before updating data on disk the actual time average is obtained by dividing for the total number of samples.

\subsubsection{Step 3}
Step 3 of the algorithm is the last one, and involves computing the wall-normal derivatives of $\aver{\delta u^2}$, which appear in the viscous parts of $\Phi_{r_y}$ and $\phi$: the viscous contributions to $\Phi_{r_y}$ and $\phi$ contain derivatives with respect to $r_y$ and $Y$ respectively. The native space where the GKE terms are computed involves $Y_1$ and $Y_2$ as independent variables, hence the following relations are used to convert derivatives between the two pairs of coordinates:
\begin{equation}
\frac{\partial}{\partial r_y}\aver{\delta u^2}=\frac{1}{2}\left( \frac{\partial}{\partial Y_2} - \frac{\partial}{\partial Y_1} \right) \aver{\delta u^2}
\end{equation}
\begin{equation}
\frac{\partial}{\partial Y}\aver{\delta u^2}=\left( \frac{\partial}{\partial Y_2} + \frac{\partial}{\partial Y_1} \right) \aver{\delta u^2}.
\end{equation}

The $Y_1$- and $Y_2$-derivatives are discretised via finite-differences, albeit not compact, with a five-points computational stencil that for the sake of consistency is identical to the one employed in the DNS code used to produce the database. Symmetries are invoked also within this step when values of $\aver{\delta u^2}$ to fill the stencil are needed in correspondence of non-available wall-normal positions $(Y_1,Y_2)$ (see Sec.\ref{sec:symmetries}).

\subsection{Exploiting symmetries}
\label{sec:symmetries}

In Ref. \cite{cimarelli-deangelis-casciola-2013} all the symmetries that characterise the terms of the GKE are comprehensively described. We take advantage of these symmetries to avoid computing the GKE terms with $Y_2<Y_1$, $Y_2>2h-Y_1$, or $Y_1>h$. 

However, these terms may be needed when wall-normal derivatives near the $Y_1$ and $Y_2$ boundaries have to be computed, and the stencil includes missing terms. These terms can be recovered by resorting to their symmetric or anti-symmetric behaviour with respect to an inversion of both $y$ (statistical symmetry) and $\bm{r}$ (analytical symmetry). Here we show how symmetries can be exploited to obtain the missing GKE terms in the $(Y_1,Y_2)$ planes with $Y_2<Y_1$, or $Y_1<h$ and $Y_2>2h-Y_1$, which is the most general case.

\begin{figure}
 \includegraphics[]{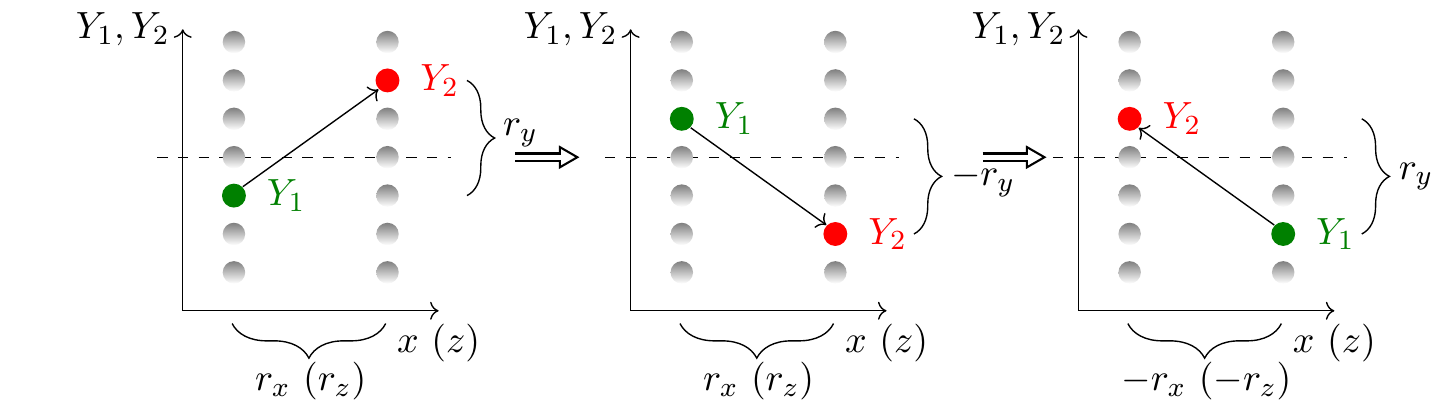}
 \caption{Graphical representation of the symmetries used to recover the GKE terms in the planes of the 4-dimensional domain $(r_x,r_z,Y_1,Y_2)$ with $Y_1<h$ and $Y_2>2h-Y_1$. The dashed line denotes the centreline of the channel, i.e. $Y_i=h$ . From the left panel to the central one we use the inversion of $y$, whereas from the central one to the left one the inversion of $\bm{r}$.}
\label{sketch-symmetry-2}
\end{figure}

We deal first with the case $Y_2<Y_1$. As shown in the two rightmost panels of figure \ref{sketch-symmetry-2}, a GKE term at a given $(r_x,r_z,Y_1,Y_2)$ with $Y_2<Y_1$ is related to an available one via the inversion of the separation vector $\vect{r}$. For example, for the scale energy we have:
\begin{equation*}
\aver{\delta u^2}(r_x,r_z,Y_1,Y_2)=\aver{\delta u^2}(-r_x,-r_z,Y_2,Y_1).
\end{equation*}
Of course the sign of a specific term must be changed according to its symmetric or anti-symmetric behaviour with respect to an inversion of $\vect{r}$. These symmetries are listed in Ref. \cite{cimarelli-deangelis-casciola-2013}.

In the case where $Y_1<h$ and $Y_2>2h-Y_1$, as shown in figure \ref{sketch-symmetry-2}, the missing GKE terms can be obtained by combining two symmetries. First the inversion of $y$ is used, arriving at an intermediate point, where the GKE terms are again not computed: taking again $\aver{\delta u^2}$ as example:
\begin{equation*}
\aver{\delta u^2}(r_x,r_z,Y_1,Y_2)=\aver{\delta u^2}(r_x,r_z,2h-Y_1,2h-Y_2).
\end{equation*}
At this point, the term although still unavailable qualifies for the previous case. Hence, as above the inversion of the separation vector can be used, arriving at available GKE terms:
\begin{equation*}
\aver{\delta u^2}(r_x,r_z,Y_1,Y_2)=\aver{\delta u^2}(-r_x,-r_z,2h-Y_2,2h-Y_1).
\end{equation*}

Again, the signs of GKE terms must be properly changed by combining the two steps. The sign of $\Phi_{r_x}$, $\Phi_{r_z}$ and $\phi$ changes, whereas the sign of $\Phi_{r_y}$, $\xi$ and $\aver{\delta u^2}$ is preserved.  

\subsection{Undersampling and parallelisation}

In Step $2$ the code allows for undersampling along the $r_x$ and $r_z$ directions, in order to reduce the memory requirement that would otherwise become excessive at high $Re$. Correlations are always computed on the full grid to exploit the Fourier representation, but when the structure $\gamma$ is assembled a smaller grid can optionally be used, which modulates the spatial resolution. The code provides for the specification of two thresholds for each of the separations $r_x$ and $r_z$. When the separation is below the first threshold, full resolution is retained; between the two thresholds a level of undersampling can be chosen, and above the second threshold an higher level of undersampling can be selected. Since undersampling is performed \emph{a posteriori}, i.e. at the time of storing the computed statistics to disk, it causes no loss of accuracy. This applies also to the accuracy of all differential operators, which are applied to the full-resolution data and are thus not altered by undersampling.

In terms of parallel computing, the code for the GKE analysis can be run serially, as it is optimised for RAM and CPU and provides for arbitrary undersampling of the data. Obviously, though, it is often convenient to exploit parallel computing. The code is equipped with three distinct parallel strategies, which can be combined at will depending upon the available computing hardware, the size of the database, and its access pattern. Note that the present implementation does not address parallelisation of the I/O operations.

First of all, a shared-memory (be it multi-core and/or multi-CPU) parallelisation is available: the user can select the number of threads to be spawned. This is particularly convenient on standard computing machines equipped with a low number of cores, for which the scaling properties in reducing the computing time for the convolutions is very good. 

In addition, the possibility exists of a domain decomposition in the wall-normal direction, so that the result space is subdivided into slices, and each slice is assigned to a different computing machine (distributed memory), inheriting what is implemented in the DNS solver \cite{luchini-quadrio-2006}. This possibility rests upon the local character of the finite-differences discretisation in the wall-normal direction. Each machine carries out independent calculations, so the parallel efficiency is the highest, although the entire velocity field must reside in the RAM of each process. Moreover, this strategy contributes to improving I/O, by employing more than one motherboard / controller / hard disk at the same time.

Lastly, the key Step 2 of the algorithm can be carried out via independent jobs, each of them dealing with only a fraction of the database. The jobs are fully independent and this strategy too trivially achieves linear scaling. There is a caveat, however: to achieve the best performance, the database must be stored in a distributed fashion. If this is not the case, there is potential for input/output contention, and the scalability of this strategy largely depends on the specific storage hardware available. Another minor drawback of this strategy is that between Step 2 and Step 3 an additional merging operation is required to bring together the various partial statistics and to carry out the final ensemble average. 

Scaling results for the first strategy will be presented in the next Section. It is noteworthy that the other two strategies have basically 100\% efficiency. Which combination of parallel strategies is best largely depends on the specific situation, the problem size and hardware availability. 

In closing, we mention that by commenting out a single line of code the user can easily switch between the version described above, where the entire snapshot is read at once and resides in RAM, and an alternate version where, for a given snapshot and a given $i_1$ plane, the $i_2$ planes are read one at a time. The first version obviously achieves a smaller I/O load, at the price of a larger RAM occupation. 
\clearpage{}
\clearpage{}\section{Results}
\label{sec:results}

We describe in this Section a typical GKE analysis, with emphasis on the computational performance of the code. We also provide some example results to highlight the novel features made possible by an efficient code, like the simultaneous access to the four-dimensional space.

\subsection{Computational details}
\label{sec:computation-details}

Two DNS databases are created for an indefinite turbulent plane channel flow at $Re_\tau=200$ and $Re_\tau=1000$. The DNS code is described in \cite{luchini-quadrio-2006}, and is a classic pseudo-spectral code with a compact, fourth-order finite-differences discretisation for the wall-normal direction and Fourier discretisation for the homogeneous directions. 

The case at $Re_\tau=200$ has $L_x= 4 \pi h$ and $L_z = 2 \pi h$, with 384 Fourier modes (256 before dealiasing) in the homogeneous directions, and 256 points in the wall-normal direction. The size of the computational domain remains unchanged for the case at $Re_\tau=1000$, while the number of modes increases to 1536, and the number of wall-normal points to 500. 

For the low-$Re$ case, the  database is made by 200 snapshots, collected at well separated times over the total duration of the simulation, i.e. about 25,000 viscous time units. The database contains the wall-normal component of the velocity and vorticity vectors, in the form of Fourier coefficients for the expansion of the variables along the homogeneous directions. The other velocity components as well as the pressure field are computed during Step 1 of the GKE analysis, as previously described in Sec.\ref{sec:structure}. The total database size is about 79 GBytes. The higher-$Re$ database is made by 35 snapshots only, but the size of the single field is larger, such that the total database size increases to 276 GBytes.

The size of the GKE database is $112$ GBytes for the low-$Re$ case, where full resolution is used. For the high-$Re$ case, the two threshold values for both the streamwise and spanwise separations are set at $200$ and $500$ in wall units. Full resolution is used below the first threshold, one every four points is retained between the thresholds, and one every eight points is retained above the second threshold. With these choices the size of the GKE database becomes $209$ GBytes at $Re_\tau=1000$.

\subsection{Code performance}
\label{sec:performances}

First, we report the outcome of a one-to-one comparison in terms of computational requirements between our code and an existing implementation, used for example by \cite{cimarelli-etal-2016} to carry out one of the most computationally demanding GKE analysis reported so far. The two codes have been re-compiled for the target machines, and tested on the same database at $Re_\tau=200$, with (384,256,384) points (no undersampling). A case with twice the number of points in every direction is also run to assess how the performance of the present solver varies with problem size. The computer where performance metrics have been measured is equipped with four AMD 6376 processors, with 16 cores each for a total of 64 cores. Clock frequency is 2.3 GHz. The I/O configuration is one of the most unfavourable, with the snapshots residing on a remote hard disk accessed via the slow NFS protocol, while output is written locally to a Western Digital 3 TBytes hard drive rotating at 5400 rpm.  

By using a single core of one CPU, i.e. in strictly scalar mode, the present code requires 239 minutes to complete the most expensive Step2 of the GKE analysis on a single flow field of the $Re_\tau=200$ database, including both I/O and CPU time. The same operation, attempted with the alternate code, takes too long for an actual measure. However, by extrapolating the time required to process a single $Y_1 - Y_2$ pair, the execution time turns out to be 3,289 times longer, i.e. about 1.5 years. Such speedup by three order of magnitudes is indeed not inconsistent with the expected speed gain when one resorts to the pseudo-spectral approach and two homogeneous directions are available on this problem size. The same test is repeated on a problem with twice the number of points in every spatial direction: the observed speedup becomes of 24,305 times, consistent with the increased size of the computational problem. Note that speedup is here defined as the ratio between the time-to-solution of the reference literature implementation and the present open-source implementation. In terms of memory requirements, our code is quite optimised, at the price of an increased I/O load. It requires 2 GBytes of RAM for the smaller case, and 18.2 GBytes for the larger case.

\begin{figure}
\centering
\includegraphics[]{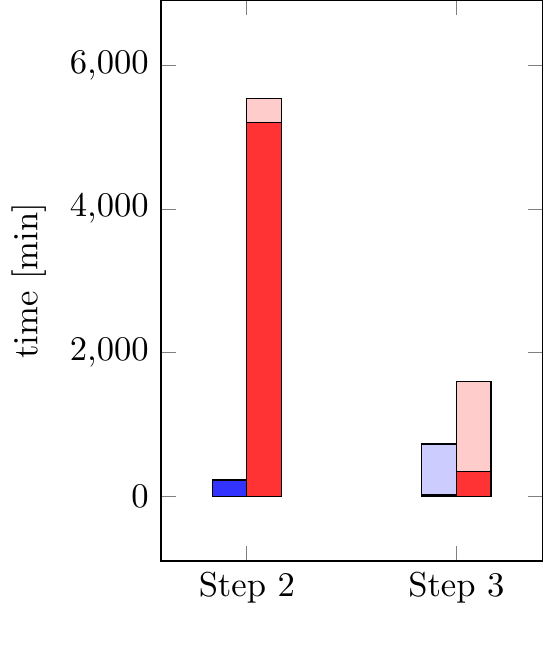}\hspace{25pt}\includegraphics[]{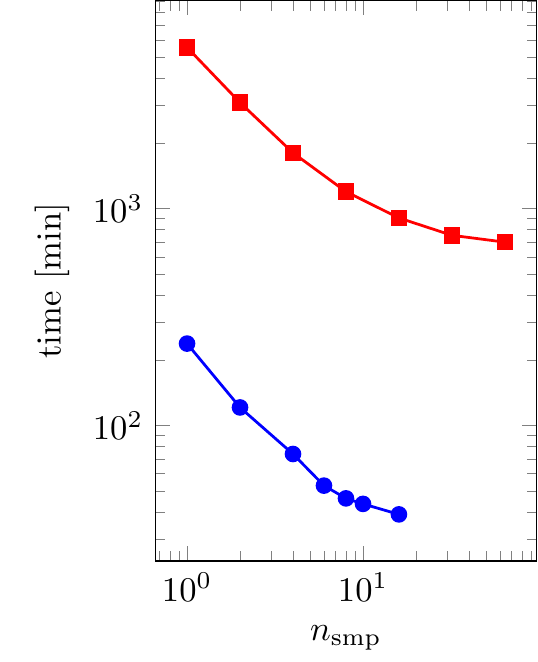}
\caption{Computational performance of the present GKE implementation. Left: wall-clock time required for Step 2 and Step 3, computed in serial mode and for a single flow field. Computing time is further divided in CPU work (dark colour) and input/output operations (bright color). Blue is the smaller case corresponding to the $Re_\tau=200$ database; red is the larger case, with twice the number of points in each spatial direction. Right: total wall-clock time required for Step 2 on a single flow field, versus the number of symmetric multiprocessing (SMP) threads $n_{smp}$. Colours as in the left panel.}
\label{fig:profiling}
\end{figure}

Figure \ref{fig:profiling} (left) further splits the computing requirements by discriminating the time needed to carry out Step 2 and Step 3. Of course, one should bear in mind that Step 2 not only is the most CPU-intensive, but also needs to be executed for as many flow fields the database is made of, whereas Step 3 needs to be executed only once. At both problem sizes, the plot shows that the program is not I/O limited, despite our architectural choice of increasing I/O load in order to alleviate memory requirements. This is remarkable, in view of the fact that I/O is quite slow on our system, and has received no optimisation at all.  Moreover, thanks to undersampling I/O is expected to only marginally increase with problem size in real use cases. I/O becomes significant only for Step 3, but this is largely expected and of no major concern, as Step 3 is a sort of post-processing step that runs only once. (Analogously, Step 1 is run only once at the pre-processing stage). 

The right panel of fig. \ref{fig:profiling} shows how the computing requirements are alleviated by the shared-memory parallel strategy. We report figures for Step 2 only, but the I/O contributions are included, and I/O is not expected to scale particularly well. Despite I/O, one observes very good speedup also on the smaller problem size and with relatively large number of cores: a single field of the $Re_\tau=200$ database can be processed by the 16 cores of a single CPU in about half an hour. This figure becomes about 13 hours for the larger test problem. It should be recalled that the other two parallel strategies mentioned earlier in Sec.\ref{sec:algorithm} possess ideal scaling properties, and that all the three available strategies can be used together to shorten the computing time. The availability of computing and storage hardware, as well as the problem size, dictate the best overall strategy on a case-by-case basis.

\subsection{GKE and turbulence physics}
\label{sec:physics}

\begin{figure}
\centering
 \includegraphics[trim=10 0 0 0]{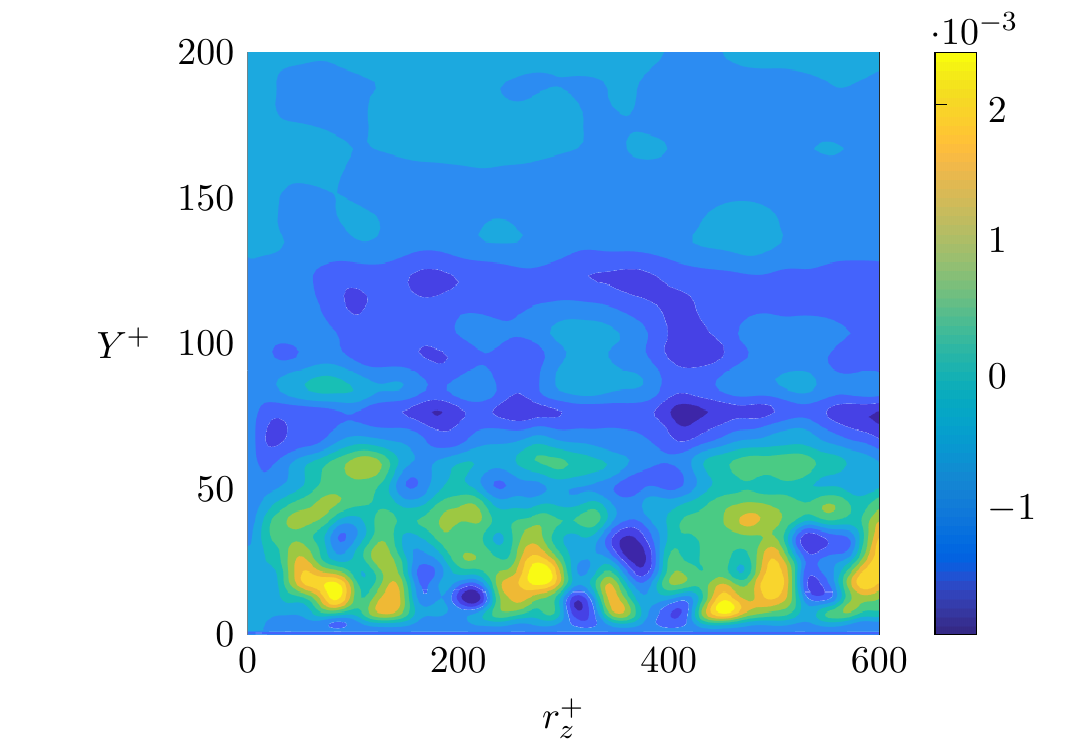}
 \caption{Residual of the GKE equation applied to the channel flow  ($Re_\tau=200$) in the $Y, r_z$ plane with $r_x^+=5$ and $r_y^+=0$. All quantities are expressed in viscous units.}
\label{fig:residuals}
\end{figure}

First, the GKE algorithm is validated by computing the residual of the budget equation \eqref{eq:GKE} in the whole 4-dimensional space, and by verifying that it is negligible everywhere in comparison to the dissipation and production terms. In doing this we verify also the statistical convergence of the data. The residual is computed with same accuracy of the GKE analysis; i.e. when computing the divergence of fluxes. The required derivatives in the homogeneous directions are performed spectrally, and those in wall-normal direction with the same high-order finite-differences scheme used elsewhere. From a quantitative point of view, the absolute maximum of the residual in the entire volume for the $Re_\tau=200$ database is $0.0104$, which is negligible when compared to the maximum and minimum of the production and dissipation terms, $1.24$ and $-0.78$ respectively. Figure \ref{fig:residuals} plots the residual of the GKE equation in the $r_x^+=5, r_y^+=0$ plane, chosen as a generic representative planar cut of the computational domain. The spatial distribution of the residual does not show any structure but one that can be attributed to remaining statistical noise, with the largest values occurring in the near-wall region where the GKE terms are also larger. In this plane, the maximum of the residual is  $0.0027$; to put this figure in perspective, the maximum and minimum of the production and dissipation terms in the same plane are $1.24$ and $-0.725$. The residual has been also verified to decrease with a larger  size of the dataset available for computing statistics.

A brief analysis of the two comprehensive newly-generated and publicly accessible GKE datasets is now presented. They illustrate the spatial and scale features of turbulent wall-bounded flows, as well as their Reynolds-number dependence. In fact, the dynamics of wall turbulence becomes richer as the Reynolds number increases; some features, absent at $Re_\tau=200$, begin to emerge at $Re_\tau=1000$. This underlines the importance of investigating high-$Re$ turbulent flows, and emphasises the need for highly efficient numerical tools.

\begin{figure}
\centering
\includegraphics[scale=0.25]{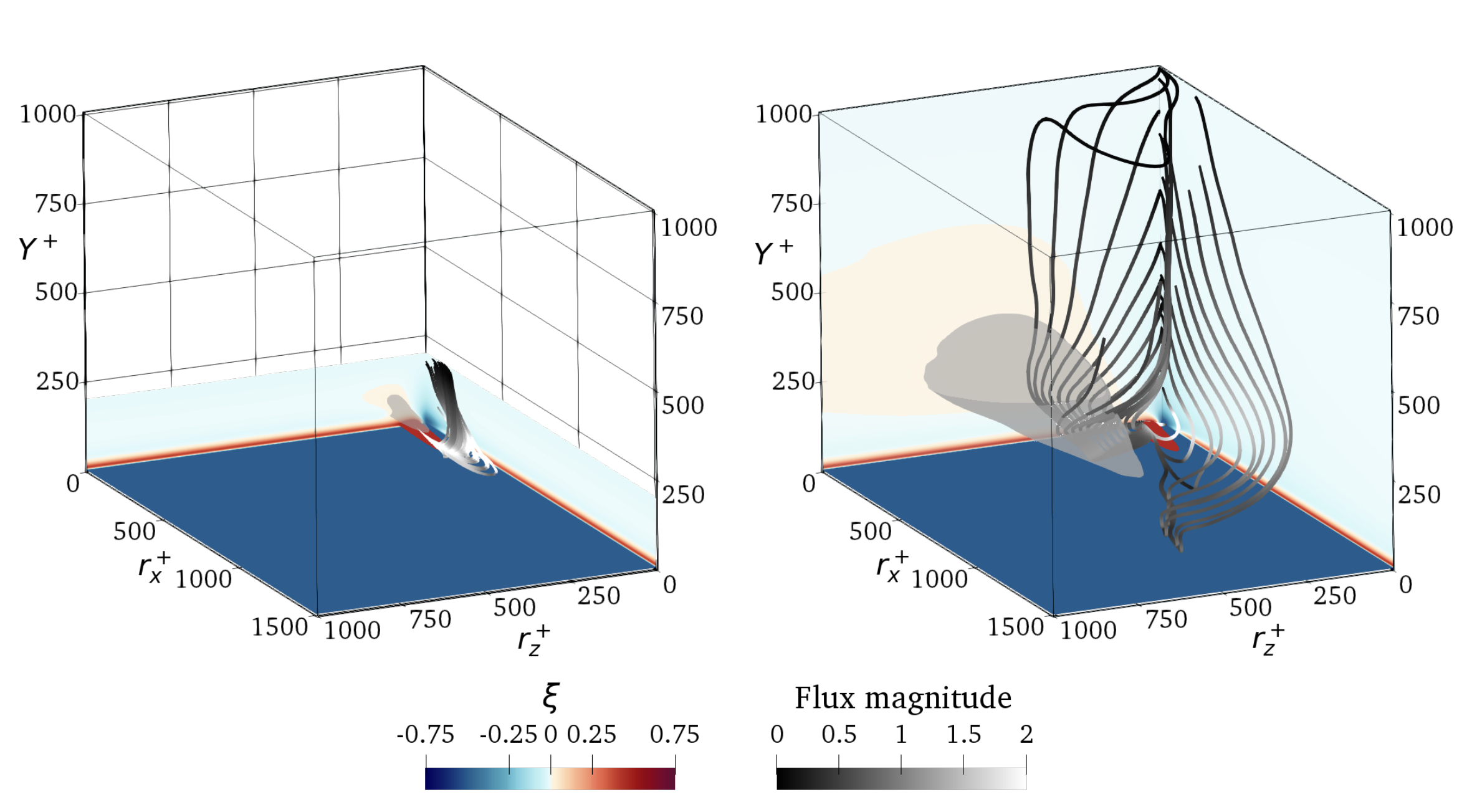}\\
\includegraphics[scale=0.25]{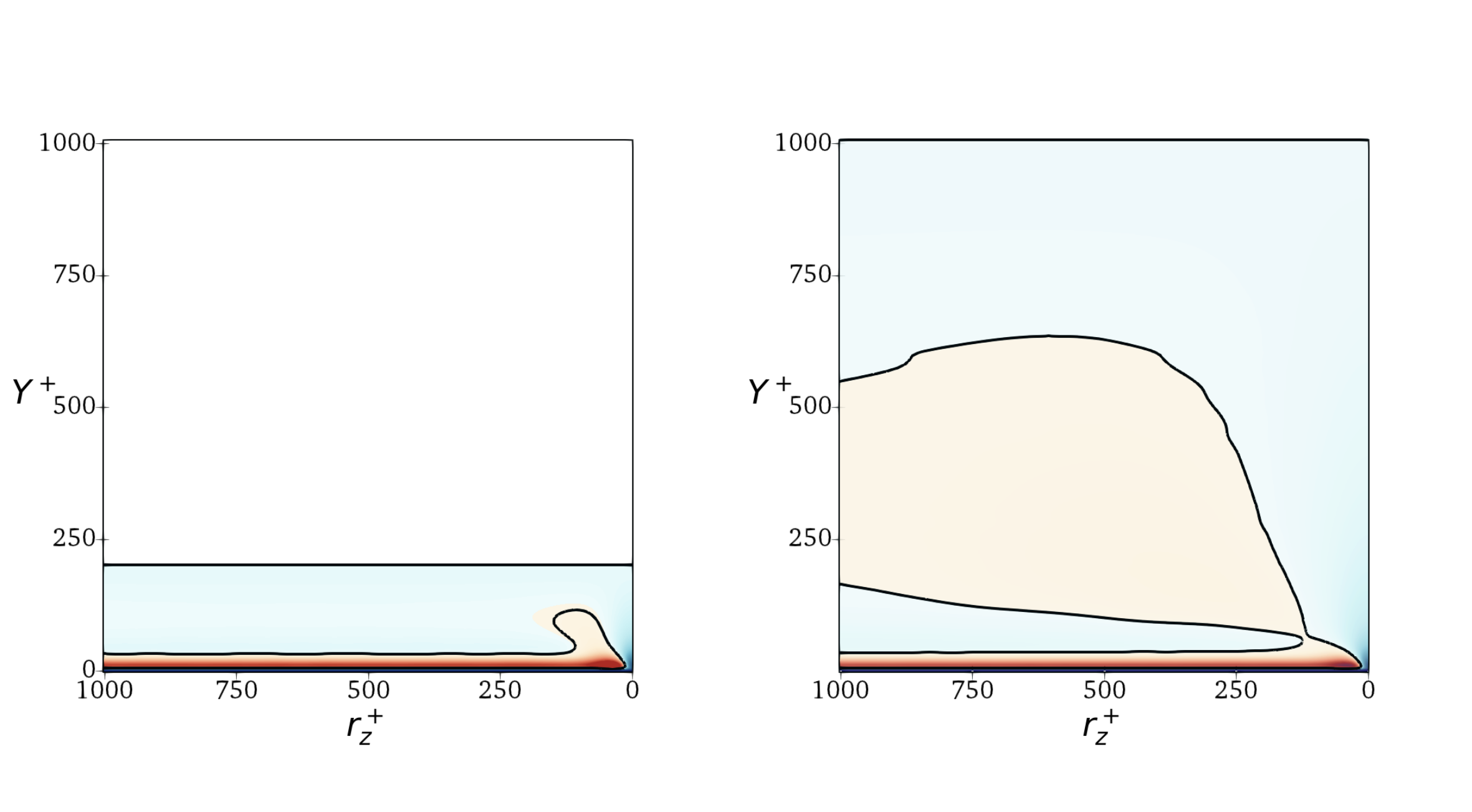}
\caption{Top: three-dimensional view of the source term $\xi$ and the vector field of fluxes $(\Phi_{r_x},\Phi_{r_z},\phi)$ in the space $r_y=0$: comparative view for $Re_\tau=200$ (left) and $Re_\tau=1000$ (right). The source field is plotted via a red isosurface corresponding to $\xi^+=0.45$, a grey isosurface corresponding to $\xi^+=0.005$, and via the two color contour planes at $r_x^+=0$ and $r_z^+=0$. The field lines are tangent to the flux vector, and are coloured according to the flux vector magnitude. Bottom: two-dimensional view of the source term $\xi^+$ in the space $r_x^+=r_y^+=0$: comparative view for $Re_\tau=200$ (left) and $Re_\tau=1000$ (right). Thick lines indicate $\xi^+=0$.}
\label{fig:3d}
\end{figure}

The GKE terms are first observed in the $r_y=0$ space. The top panels of figure \ref{fig:3d} feature the source term $\xi$ and the fluxes $(\Phi_{r_x},\Phi_{r_z},\phi)$ in this 3-dimensional space, comparing the $Re_\tau=200$ case on the left to the $Re_\tau=1000$ case on the right; the bottom panels plot a two-dimensional section of the volume taken at $r_x=0$. The figures use the same scale on the axes, so that the effects of increasing the Reynolds number can be easily appreciated.

The emerging picture, already described for example in Ref. \cite{cimarelli-deangelis-casciola-2013}, is that in both cases near the wall a region with large positive $\xi$ is present, where energy production largely overcomes its dissipation rate; see the red isosurfaces (corresponding to $\xi^+=0.45$) visible at small scales and wall distances in the top panels of figure \ref{fig:3d} and the near-wall peak of the contour in the bottom panels. This is where wall turbulence is mainly produced. The extent of this region scales in wall units, hence it shrinks in absolute terms with increasing $Re$. The scales $0<r_x^+<200$, $25<r_z^+<70$ and $Y^+ \sim 13$ shown by the red isosurface, suggest a strong connection with the main coherent structures in the wall region: the quasi-streamwise vortices and the streaks of streamwise velocity \cite{jeong-etal-1997, robinson-1991b}. On the other hand, large negative values of the source term are observed at $Y \rightarrow 0$ for any scale, and at $r_x \rightarrow 0$ and $r_z \rightarrow 0$ for any wall distance. Accordingly, the immediate vicinity of the wall and the smallest scales of motion in the whole flow are recognised to be the sink regions of wall turbulence, where viscous dissipation dominates.

Only in the high-$Re$ case, a further large region of positive $\xi$ is additionally seen quite far from the wall, in correspondence of larger streamwise and spanwise scales, separated from the near-wall peak by a (sink) region with $\xi<0$. This is in agreement with the results shown in \cite{cimarelli-etal-2015} using different DNS databases at $Re_{\tau}=550,1000,1500$. This region, absent in the low-$Re$ case, presents rather low values of $\xi$, about one order of magnitude smaller than the values of the near-wall production region, with the peak of value $\xi^+=0.0095$ placed at $r_x^+=0$, $r_z^+ \sim 350$ and $Y^+ \sim 160$ (same findings of Ref. \cite{cimarelli-etal-2015}). This secondary peak of the source term is related to an outer self-sustained mechanism of turbulence well separated from the near-wall dynamics, as discussed in Refs. \cite{hutchins-marusic-2007, jimenez-hoyas-2008, smits-mckeon-marusic-2011} and several others. The scales and wall distances at which it occurs are in agreement with the findings of Ref. \cite{hutchins-marusic-2007}. Of course, at $Re_\tau=1000$ the outer peak is only beginning to appear, and the two peaks do not show yet a complete separation: see the contours of the source term in the bottom right panel. Refs. \cite{hutchins-marusic-2007} and \cite{smits-mckeon-marusic-2011}, by observing one-dimensional premultiplied power spectra of $\aver{u^2}$ at progressively higher Reynolds numbers, suggest that $Re_\tau$ approximately larger than $1700$ is required before the outer site can be clearly noticed. Since our data show an outer peak already at $Re_\tau=1000$, it is possible that the GKE provides an earlier and/or sharper detection of the outer cycle compared to the premultiplied spectra, as already hypothesised by \cite{davidson-nickels-krogstad-2006} and, more recently, by \cite{agostini-leschziner-2017}, where the differences in the detection of the $k^{-1}$ spectral law and of the real-space analogue logarithmic dependence on $r_x$ of the streamwise structure function are discussed.

The GKE also provides us with the knowledge of the field of energy fluxes. This can be exploited, along the lines of Refs. \cite{cimarelli-deangelis-casciola-2013} and \cite{cimarelli-etal-2016}, to follow scale energy as it moves from the source regions to the sink regions, tracking the involved scales and wall distances. This is visualised in the $r_y=0$ space of figure \ref{fig:3d} by field lines tangent to the flux vector $(\Phi_{r_x},\Phi_{r_z}, \phi)$. In both the low-$Re$ and high-$Re$ cases, the flux lines origin from a singularity point located close to the peak of the source term in the near-wall region, i.e. $r_x^+=0$, $r_z^+ \sim 60$ and $Y^+ \sim 14$, and are attracted by the two sink regions mentioned above (the wall plane, and the $r_x=r_z=0$ axis at larger wall distances). From a topological viewpoint, the lines fulfil the requirement \cite{cimarelli-deangelis-casciola-2013} of vanishing perpendicularly to the sinks. In accordance with the outcome of the single-point budget for the turbulent kinetic energy, these lines reveal that the excess of turbulent energy production in the buffer layer feeds both the upper and the lower regions. However, the GKE provides important additional information concerning the scales involved in these spatial transfers. For example, following the lines of the branch vanishing at large $Y$, the coexistence of reverse and direct cascades is observed while turbulent energy ascends from the wall. In detail, as shown in figure \ref{fig:3d}, the lines emanating from the singularity point show first an inverse cascade of energy moving towards increasing $r_x$ and $r_z$. Later a mixed direct/inverse cascade takes place, while $Y$ remains constant: an inverse cascade towards increasing $r_x$ is seen together with a direct cascade towards decreasing $r_z$. Finally, the lines start ascending towards larger $Y$ and present a more classic direct cascade towards smaller $r_x$ and $r_z$. Interestingly, in the  $Re_\tau=1000$ case some of the lines that feed the sink at larger wall distances, are seen to cross the outer source peak; they feed the field with the excess production associated with the large-scale outer motions.

\begin{figure}
\centering
\includegraphics[scale=0.27]{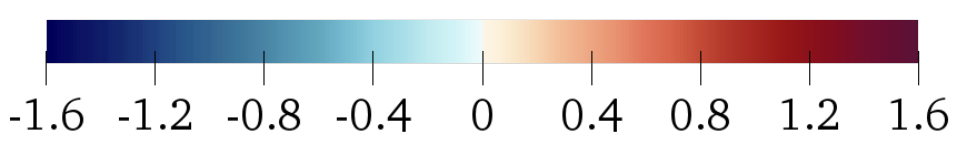}\\
\includegraphics[scale=0.23]{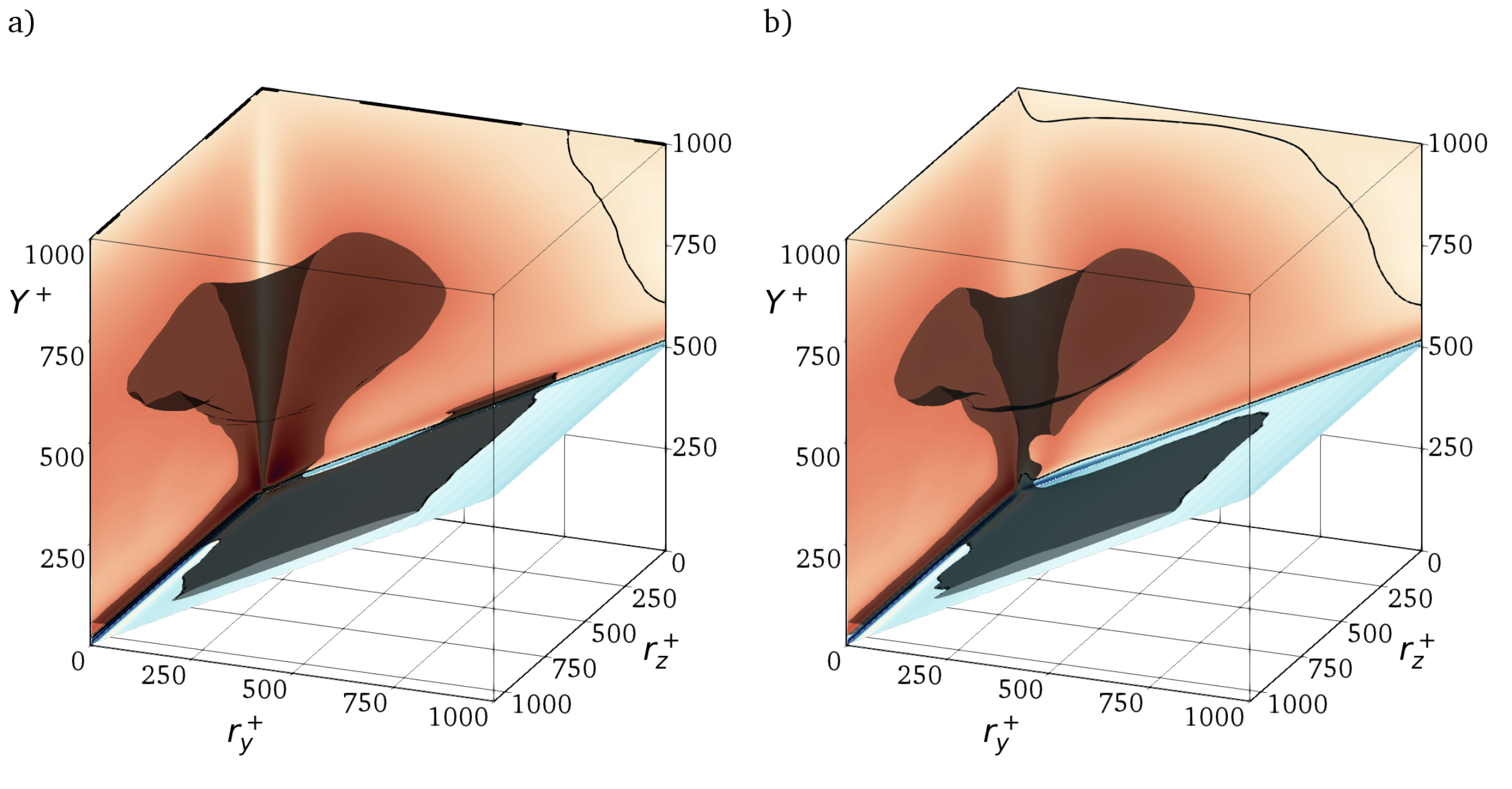}\\
\includegraphics[scale=0.23]{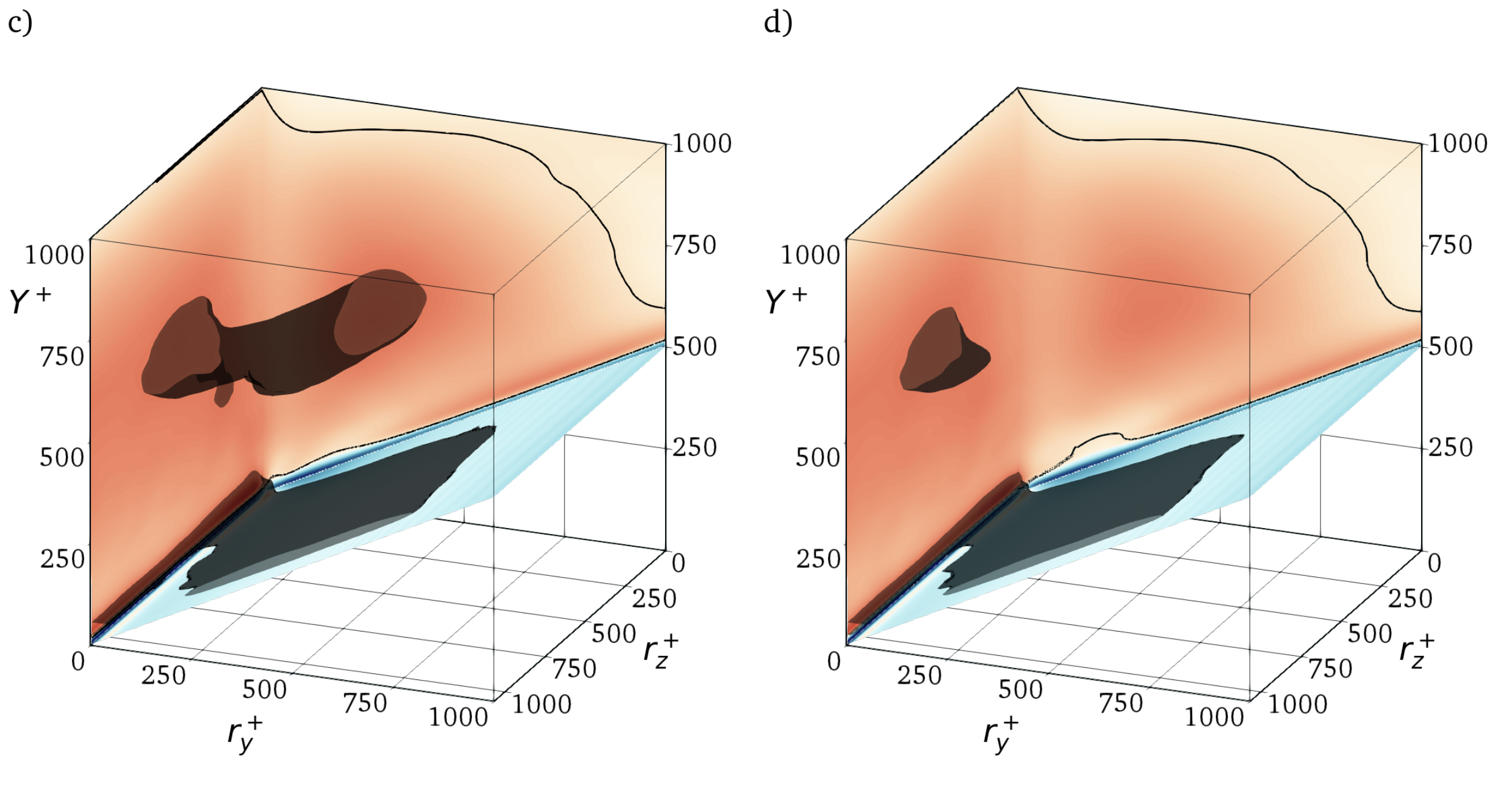}\\
\includegraphics[scale=0.23]{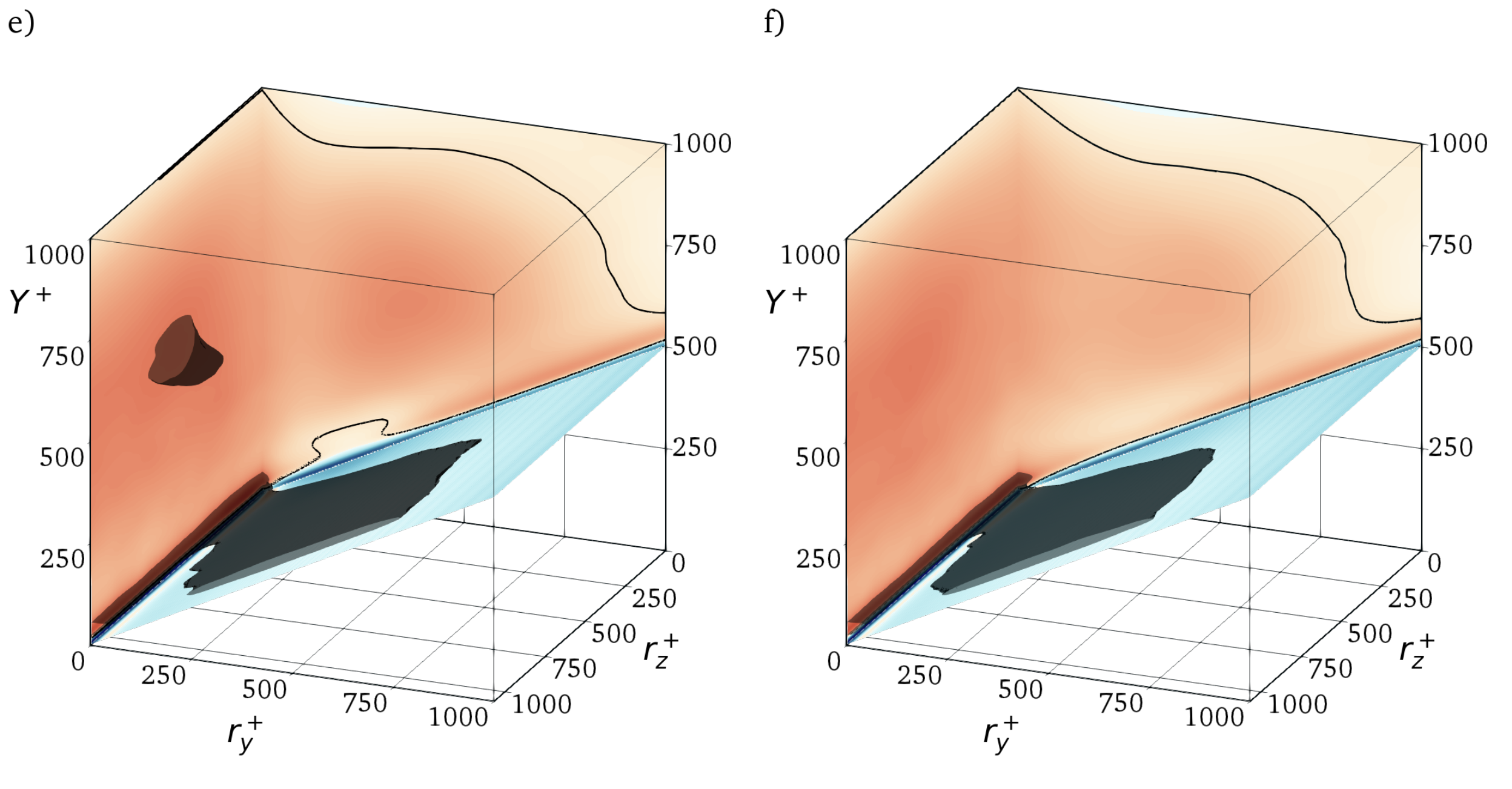}\\
\caption{Plot of $\phi$ in the $r_x^+=0$ (a), $r_x^+=50$ (b), $r_x^+=100$ (c), $r_x^+=150$ (d), $r_x^+=200$ (e) and $r_x^+=400$ (f) 3-dimensional spaces. Dark-coloured isosurface corresponds to $\phi=0.5$. Black isolines at the $r_z=0$ and $r_y=0$ planes correspond to $\phi=0$. The accompanying movie can be found as supplemental material to this manuscript.}
\label{fig:4d}
\end{figure}

The efficient implementation of the GKE analysis makes it affordable to look at the 4-dimensional domain overall. Figure \ref{fig:4d} plots six 3-dimensional volumes, extracted at different streamwise separations from a 4-dimensional dataset. Moreover, the full dataset is shown in the supplemental material to this manuscript as a movie, where $r_x$ is used as the temporal dimension to build the animation. The GKE dataset underlying the Figure can be freely downloaded at this \href{https://github.com/davecats/gke/tree/master/database}{link}. The separations extracted to produce figure \ref{fig:4d} are $r_x^+=0, 50, 100, 150, 200, 400$, whereas in the movie the separation varies continuously from $r_x^+=0$ to $r_x^+=1000$. In the 3-dimensional space where $r_x$ is fixed, i.e. on a frame of the movie, the GKE terms are not defined below the $Y={r_y}/2$ plane, owing to the finite size of the channel in the wall-normal direction. 

The plotted quantity is the wall-normal flux $\phi^+$ as it is the one changing the most along the $r_x$ direction. It is represented via contour planes as well as with a dark-coloured isosurface corresponding to the value $\phi^+=0.5$ and black isolines corresponding to $\phi^+=0$. Note that at $r_x=0$, the largest values of $\phi$ are seen in the near-wall region at small $r_y$. The maximum of about 1.5 is observed at zero spanwise separations, namely at $(r_y^+,r_z^+,Y^+) \sim (50,0,30)$. This maximum is associated with the near-wall cycle, as these separations and wall-normal positions are consistent with the findings of Ref. \cite{jeong-etal-1997} concerning the dominant near-wall vortical structures. At larger $r_x$, as shown in the other panels, the near-wall maximum is nearly unchanged, except for the portion near $r_z=0$, i.e. the statistical footprint of the near-wall cycle, which decreases. On the contrary, the largest negative values are always for $r_y \rightarrow 0$ and $Y^+<20$. Accordingly, for the budget equation of the turbulent kinetic energy, which is recovered here for $r_y=0$, $r_z \rightarrow L_z/2$ and $r_x \rightarrow L_x/2$, the flux shows a non-monotonic behaviour with negative values for $Y^+<17$ and a positive peak placed at $Y^+=37$, in agreement with results presented in Refs. \cite{marati-casciola-piva-2004} at $Re_{\tau}=180$ and \cite{cimarelli-etal-2015} at $Re_{\tau}=550,1000,1500$.

Large positive values of $\phi$ are also observed in a flat region in the vicinity of the $Y^+={r_y^+}/2+30$ plane for $r_y^+<750$ and $r_z^+>200$, excluding the smallest wall-normal and spanwise scales. In the $r_x=0$ volume, unlike at larger streamwise separations, this region is observed to connect to the $r_z=0$ plane, but this connection is lost at larger $r_x$. Since along the oblique plane the wall-normal positions of the points used to compute the velocity increments are $y_1^+=30$ and $y_2^+=30+r_y^+$, a large positive value of the spatial flux implies that attached eddies in the sense of Ref. \cite{townsend-1976} are associated with an outward flux of $\aver{\delta u^2}$ \cite{cimarelli-etal-2016}.

In the $r_x=0$ volume, negative values of $\phi$ are seen only very close to the $Y={r_y}/2$ plane and for large $Y$ and $r_y$ (see the black contour line in the $r_z=0$ plane in panel (a), denoting the zero level). At increasing $r_x$ an additional region with negative $\phi$ is found at $r_z \rightarrow 0$, $Y^+ \sim 180 $ and $r_y^+ \sim 250$. Interestingly, this region reaches its largest extension for $r_x^+=200$ before disappearing with increasing streamwise separation.

Finally, for $r_x^+ \le 200$, large values of $\phi$ are also observed at $Y^+ \sim 500$, i.e. in the log-layer. In detail, at $r_x=0$, $\phi^+$ is larger than $0.5$ for $r_z^+<500$ and $r_y^+300$, excluding the smallest scales. By increasing $r_x$, $\phi$ decreases in the logarithmic layer. Interestingly, the decrease rate is faster at small $r_z$. In fact, large values of $\phi$ are still present at $r_y \rightarrow 0$ and $r_z^+ \sim 700$ in the volumes with $r_x^+=150$ and $r_x^+=200$. The large values of $\phi$ observed in the log-layer are associated with the structures of the outer self-sustained mechanism of turbulence, as these $r_z$ and $Y$ are in agreement with the findings of Ref. \cite{hutchins-marusic-2007}. Hence, a large positive spatial flux of $\aver{\delta u^2}$ may be related to these large-scale motions. The region with large $\phi$ placed near the wall is not separated from the one in the log-layer in the volumes with $r_x^+=0$ and $r_x^+=50$. On the contrary, for $r_x^+ > 50 $, these two regions of large $\phi$ are not connected, as shown by the isosurface in panel (c) and (d), denoting a separation between the phenomena traceable to the near-wall structures and those to the outer structures.

\clearpage{}

\section{Conclusions and outlook}

This work has described the implementation of a parallel computer program that builds the complete budget of the Generalised Kolmogorov equation (GKE) starting from a DNS-produced database of a turbulent channel flow. The source code is freely available on GitHub. The most important feature is that the terms of the GKE, made by products of velocity and pressure differences, are rewritten as sums of cross-correlations. When homogeneous directions are present (the indefinite plane channel flow possesses two of them), the Parseval theorem allows efficient computation of such correlations in Fourier space, with huge computational advantages. These advantages become more and more significant as the size of the computational problem increases, as it is expected when dealing with high-$Re$ flows; they also remain significant when the homogeneous direction is only one, thus providing the present approach with a much broader scope than the indefinite plane channel flow alone consdered in the present work. 

Several optimisations are used to keep the CPU and RAM requirements to a minimum. Extensive use of analytical and statistical symmetries reduces the number of functional evaluations required to compute all the terms in the whole four-dimensional space of their independent variables. As a result, in serial mode the code has been measured to provide three- or four-orders of magnitude speedup (depending on the problem size) when compared to a standard implementation. Three distinct parallel strategies are available and can be combined freely to best match the specific hardware configuration (number and type of machines, CPU cores, storage system, etc). 

The unbalance of the GKE terms, which descends from the finite size of the statistical sample, is presented for validation; it is found to be negligible and to decrease with sample size. For the first time the complete set of terms in the GKE has been computed and observed in the whole 4-dimensional space. Results are presented for two channel flow cases, at $Re_\tau=200$ and $Re_\tau=1000$. It is shown that the present code can handle very large datasets with a reasonable amount of computational resources. Although the 5-fold separation of the considered $Re$ is limited, and $Re_\tau=1000$ can hardly be considered sufficient to achieve a turbulent flow with a well-developed outer cycle, our analysis reveals quite clearly the distinction between the inner and outer turbulence cycles, as well as the distinction between attached and detached turbulent structures. The possibility thus exists that the GKE is an effective tool to put these distinctions into focus.

The present methodology takes full advantage of the double statistical homogeneity of parallel indefinite flows (Poiseuille, Couette), but it can be readily extended to flows with two inhomogenous direction. For such flows, the analysis of the spatial and scale transfer phenomena becomes even more challenging and revealing, as recently demonstrated by \cite{mollicone-etal-2018} for turbulent flows undergoing separation. The additional inhomogeneous direction is dealt with as the wall-normal direction in the Poiseuille flow, with only straightforward modifications to the source code.

We hope that the computational tool described in this paper will enable advancing our understanding of turbulent flows. An extension of the GKE equation to deal with the anisotropic case by considering every component of the Reynolds' stress tensor is underway, which is deemed to bring even more insight into the physics of a geometrically simple but highly anisotropic flow like the channel flow. At the same time, the GKE analysis is being used to understand the profound modifications induced in a  natural turbulent flow by skin-friction drag reduction techniques. For both these goals, the availability of a reliable, efficient and compact code to carry out the GKE analysis is a crucial step towards the understanding of the complex physical processes which regulate production, transfer and dissipation of turbulent energy in a wall-bounded flow.

\section{Acknowledgements}
This work is supported by the Priority Programme SPP 1881 Turbulent Superstructures of the Deutsche Forschungsgemeinschaft. Computing time has been provided by the computational resource ForHLR Phase I funded by the Ministry of Science, Research and the Arts, Baden-W\"urttemberg and DFG.


\begin{thebibliography}{10}
\providecommand{\url}[1]{\normalfont{#1}}
\providecommand{\urlprefix}{Available from: }

\bibitem{richardson-1922}
Richardson~L. Weather prediction by numerical process. Cambridge University
  Press; 1922.

\bibitem{kolmogorov-1941}
Kolmogorov~A. The {L}ocal {S}tructure of {T}urbulence in an {I}ncompressible
  {V}iscous {F}luid for {V}ery {L}arge {R}eynolds {N}umbers. Dokl Akad Nauk
  SSSR. 1941;\hspace{0pt}30:301--305. (Reprinted in Proc. R. Soc. London A
  v.434 pp.9--13, 1991).

\bibitem{kim-moin-moser-1987}
Kim~J, Moin~P, Moser~R. Turbulence statistics in fully developed channel flow
  at low {R}eynolds number. J Fluid Mech. 1987;\hspace{0pt}177:133--166.

\bibitem{mansour-kim-moin-1988}
Mansour~N, Kim~J, Moin~P. Reynolds-stress and dissipation-rate budgets in a
  turbulent channel flow. J Fluid Mech. 1988;\hspace{0pt}194:15--44.

\bibitem{kolmogorov-dissipation-1941}
Kolmogorov~A. Dissipation of energy in locally isotropic turbulence. In: Dokl.
  Akad. Nauk SSSR; Vol.~32; 1941. p. 16--18.

\bibitem{hill-2001}
Hill~R. Equations relating structure functions of all orders. J Fluid Mech.
  2001;\hspace{0pt}434:379--388.

\bibitem{hill-2002}
Hill~R. Exact second-order structure-function relationships. J Fluid Mech.
  2002;\hspace{0pt}468:317--326.

\bibitem{danaila-etal-2001}
Danaila~L, Anselmet~F, Zhou~T, et~al. Turbulent energy scale budget equations
  in a fully developed channel flow. J Fluid Mech.
  2001;\hspace{0pt}430:87--109.

\bibitem{marati-casciola-piva-2004}
Marati~N, Casciola~C, Piva~R. Energy cascade and spatial fluxes in wall
  turbulence. J Fluid Mech. 2004;\hspace{0pt}521:191--215.

\bibitem{cimarelli-deangelis-casciola-2013}
Cimarelli~A, De~Angelis~E, Casciola~C. Paths of energy in turbulent channel
  flows. J Fluid Mech. 2013;\hspace{0pt}715:436--451.

\bibitem{cimarelli-etal-2016}
Cimarelli~A, De~Angelis~E, Jimenez~J, et~al. Cascades and wall-normal fluxes in
  turbulent channel flows. J Fluid Mech. 2016;\hspace{0pt}796:417--436.

\bibitem{cimarelli-deangelis-2011}
Cimarelli~A, De~Angelis~E. Analysis of the {K}olmogorov equation for filtered
  wall-turbulent flows. J Fluid Mech. 2011;\hspace{0pt}676:376--395.

\bibitem{cimarelli-deangelis-2012}
Cimarelli~A, De~Angelis~E. Anisotropic dynamics and sub-grid energy transfer in
  wall-turbulence. Phys Fluids. 2012;\hspace{0pt}24(1):015102.

\bibitem{gomes-ganapathisubramani-vassilicos-2015}
Gomes-Fernandes~R, Ganapathisubramani~B, Vassilicos~J. The energy cascade in
  near-field non-homogeneous non-isotropic turbulence. J Fluid Mech.
  2015;\hspace{0pt}771:676--705.

\bibitem{portela-papadakis-vassilicos-2017}
Portela~FA, Papadakis~G, Vassilicos~J. The turbulence cascade in the near wake
  of a square prism. J Fluid Mech. 2017;\hspace{0pt}825:315--352.

\bibitem{burattini-antonia-danaila-2005}
Burattini~P, Antonia~R, Danaila~L. Scale-by-scale energy budget on the axis of
  a turbulent round jet. J Turbulence. 2005;\hspace{0pt}(6):N19.

\bibitem{thiesset-antonia-danaila-2013}
Thiesset~F, Antonia~R, Danaila~L. Scale-by-scale turbulent energy budget in the
  intermediate wake of two-dimensional generators. Phys Fluids.
  2013;\hspace{0pt}25(11):115105.

\bibitem{danaila-etal-2012}
Danaila~L, Krawczynski~J, Thiesset~F, et~al. Yaglom-like equation in
  axisymmetric anisotropic turbulence. Physica D: Nonlinear Phenomena.
  2012;\hspace{0pt}241(3):216--223.

\bibitem{gauding-etal-2014}
Gauding~M, Wick~A, Pitsch~H, et~al. Generalised scale-by-scale energy-budget
  equations and large-eddy simulations of anisotropic scalar turbulence at
  various schmidt numbers. J Turbulence. 2014;\hspace{0pt}15(12):857--882.

\bibitem{togni-cimarelli-deangelis-2015}
Togni~R, Cimarelli~A, De~Angelis~E. Physical and scale-by-scale analysis of
  {R}ayleigh--{B}\'enard convection. J Fluid Mech.
  2015;\hspace{0pt}782:380--404.

\bibitem{davidson-2004}
Davidson~P. Turbulence: {A}n {I}ntroduction for {S}cientists and {E}ngineers.
  Oxford University Press; 2004.

\bibitem{germano-2007}
Germano~M. The elementary energy transfer between the two-point velocity mean
  and difference. Phys Fluids. 2007;\hspace{0pt}19(8):085105.

\bibitem{jimenez-2016}
Jim\'enez~J. Optimal fluxes and {R}eynolds stresses. J Fluid Mech.
  2016;\hspace{0pt}809:585--600.

\bibitem{luchini-quadrio-2006}
Luchini~P, Quadrio~M. A low-cost parallel implementation of direct numerical
  simulation of wall turbulence. J Comp Phys. 2006;\hspace{0pt}211(2):551--571.

\bibitem{jeong-etal-1997}
Jeong~J, Hussain~F, Schoppa~W, et~al. Coherent structures near the wall in a
  turbulent channel flow. J Fluid Mech. 1997;\hspace{0pt}332:185--214.

\bibitem{robinson-1991b}
Robinson~SK. Coherent motions in the turbulent boundary layer. Ann Rev Fluid
  Mech. 1991;\hspace{0pt}23:601--639.

\bibitem{cimarelli-etal-2015}
Cimarelli~A, De~Angelis~E, Schlatter~P, et~al. Sources and fluxes of scale
  energy in the overlap layer of wall turbulence. J Fluid Mech.
  2015;\hspace{0pt}771:407--423.

\bibitem{hutchins-marusic-2007}
Hutchins~N, Marusic~I. Evidence of very long meandering features in the
  logarithmic region of turbulent boundary layers. J Fluid Mech.
  2007;\hspace{0pt}.

\bibitem{jimenez-hoyas-2008}
Jim\'enez~J, Hoyas~S. Turbulent fluctuations above the buffer layer of
  wall-bounded flows. J Fluid Mech. 2008;\hspace{0pt}611:215--236.

\bibitem{smits-mckeon-marusic-2011}
Smits~AJ, McKeon~BJ, Marusic~I. High-{R}eynolds number wall turbulence. Annu
  Rev Fluid Mech. 2011;\hspace{0pt}43(1):353--375.

\bibitem{davidson-nickels-krogstad-2006}
Davidson~P, Nickels~T, Krogstad~PA. The logarithmic structure function law in
  wall-layer turbulence. J Fluid Mech. 2006;\hspace{0pt}550:51--60.

\bibitem{agostini-leschziner-2017}
Agostini~L, Leschziner~M. Spectral analysis of near-wall turbulence in channel
  flow at ${R}e_\tau=4200$ with emphasis on the attached-eddy hypothesis. Phys
  Rev Fluids. 2017;\hspace{0pt}2(1):014603.

\bibitem{townsend-1976}
Townsend~A. The {S}tructure {O}f {T}urbulent {S}hear {F}lows. 2nd ed. Cambridge
  University Press; 1976.

\bibitem{mollicone-etal-2018}
Mollicone~JP, Battista~F, Gualtieri~P, et~al. Turbulence dynamics in separated
  flows: the generalised kolmogorov equation for inhomogeneous anisotropic
  conditions. J Fluid Mech. 2018;\hspace{0pt}841:1012--1039.

\end{thebibliography}
\end{document}